\documentclass[3p,a4paper,times,final]{elsarticle}
\usepackage[utf8]{inputenc}
\usepackage{graphicx,xspace}
\usepackage{amsmath}
\usepackage{amssymb}
\usepackage[dvipsnames]{xcolor}
\usepackage{pgf}
\usepackage[normalem]{ulem}
\usepackage{tikz}
\usetikzlibrary{decorations,arrows}
\usetikzlibrary{decorations.pathmorphing}
\usepackage[british]{babel}
\usepackage[colorlinks=true]{hyperref}

\newcommand\definesymb[1]{%
\expandafter\newcommand\csname #1#1\endcsname{{\ensuremath{\mathbb{#1}}}}%
}

\newcommand\north{\textsf{North}}
\newcommand\south{\textsf{South}}
\newcommand\east{\textsf{East}}
\newcommand\west{\textsf{West}}

\definesymb{Z}
\definesymb{N}
\definesymb{R}

\newcommand\correc[1]{{#1}}

\newcommand\cR{}%\color{red}}
\newcommand\dff{}%\color{red}}

\newcommand\lang{\mathcal L}

\newtheorem{thm}{Theorem}
\newtheorem{cor}[thm]{Corollary}
\newtheorem{prop}[thm]{Proposition}
\newtheorem{lem}[thm]{Lemma}

\newtheorem{defn}{Definition}
\newdefinition{example}{Example}
\newproof{pf}{Proof}

\begin{document}

\begin{frontmatter}
\title{Subshifts as Models for MSO Logic\tnoteref{t1}}
\tnotetext[t1]{The authors are partially supported by ANR-09-BLAN-0164.}
\author[1]{Emmanuel Jeandel} \ead{emmanuel.jeandel@loria.fr}
\author[2]{Guillaume Theyssier}\ead{guillaume.theyssier@univ-savoie.fr}

\address[1]{LORIA, UMR 7503 - Campus Scientifique, BP 239\\
54\,506 VANDOEUVRE-LÈS-NANCY, FRANCE}
\address[2]{LAMA (Universit\'e de Savoie, CNRS)\\ Campus
Scientifique, 73376 Le Bourget-du-lac cedex FRANCE}%

%\maketitle

\begin{abstract}
  We study the Monadic Second Order (MSO) Hierarchy over colourings of
  the discrete plane, and draw links between classes of formula and
  classes of subshifts. We give a characterization of existential MSO
  in terms of projections of tilings, and of universal sentences in
  terms of combinations of ``pattern counting'' subshifts. Conversely,
  we characterise logic fragments corresponding to various classes of
  subshifts (subshifts of finite type, sofic subshifts, all
  subshifts). Finally, we show by a separation result how the
  situation here is different from the case of tiling pictures studied
  earlier by Giammarresi et al.
\end{abstract}
\end{frontmatter}
\fboxsep=3pt
\newcommand\pred[2]{P_{\fcolorbox{gray}{#1}{}} (#2)}
\newcommand\tcolor[1]{\fcolorbox{gray}{#1}{\;\vbox{\hrule height 4pt width 0pt}}}

\section{Introduction}
\label{sec:intro}
There is a close connection between words and monadic second-order (MSO) logic.
B\"uchi and Elgot proved for finite words that MSO-formulas correspond
exactly to regular languages. This relationship was developed for other
classes of labeled graphs; trees or infinite words enjoy a similar connection. 
See \cite{Thomas,Matz} for a survey of existing results. Colorings of
the entire plane, i.e tilings, represent a natural generalization of
biinfinite words to higher dimensions, and as such enjoy similar properties.
We plan to study in this paper tilings for the point of view of monadic second-order logic.

\correc{From a computer science point of view, tilings and more
  generally subshifts are the underlying objects of several computing
  models including cellular automata \cite{D11,BoyerT10,BoyerT09},
  Wang tiles \cite{LafitteW07,LafitteW08} and self-assembly tilings
  \cite{DotyPRSS10,DotyLPSW10}. Following the recent trend to better
  understand such 'natural computing models', one of the motivations
  of the present paper is to extend towards these models the fruitful
  links established between languages of finite words and MSO logic.}

Tilings and logic have a shared history. The introduction of tilings can be
traced back to Hao Wang \cite{wangpatternrecoII}, who introduced
his celebrated tiles to study the (un)decidability of
the $\forall \exists \forall$ fragment of first order logic.
The undecidability of the domino problem by his PhD Student Berger
\cite{Ber-undecidability-dp} lead then to the undecidability of this fragment
\cite{classicaldecisionproblem}. Seese \cite{Kuske,Seese} used the domino problem
to prove that graphs with a decidable MSO theory have a bounded tree width.
Makowsky\cite{makowsky74,poizat80} used the construction by Robinson \cite{Robinson}
to give the first example of a finitely axiomatizable theory that is
super-stable. More recently, Oger \cite{oger04} gave generalizations of classical results
on tilings to locally finite relational structures. See the survey
\cite{jac:modth} for more details.

Previously, a finite variant of tilings, called tiling pictures, was studied
\cite{GiamRest2,GiamRest}. Tiling pictures correspond to colorings of a \emph{finite}
region of the plane, this region being bordered by special `\texttt{\#}' symbols.
It is proven for this particular model that language recognized by 
EMSO-formulas correspond exactly to so-called finite tiling systems, 
i.e. projections of finite tilings.

The equivalent of finite tiling systems for infinite pictures are
so-called \emph{sofic subshifts} \cite{Weiss}. A \emph{sofic subshift}
represents intuitively local properties and ensures that every point
of the plane behaves in the same way.  As a consequence, there is no
general way to enforce that some specific color, say \textbf{A},
appears at least once.  Hence, some simple first-order existential
formulas have no equivalent as sofic subshift (and even
subshift). This is where the border of \texttt{\#} for finite pictures
play an important role: Without such a border, results on finite
pictures would also stumble on this issue. See \cite{AnselmoJM09} for similar
results on finite pictures without borders.

We deal primarily in this article with subshifts. See \cite{Alten}
for other acceptance conditions (what we called subshifts of finite type
correspond to A-acceptance in this paper).

Finally, note that all decision problems in our context are non-trivial : To
decide if a universal first-order formula is satisfiable (the domino
problem, presented earlier) is not recursive.
Worse, it is $\Sigma_1^1$-hard to decide if a tiling of the plane exists
where some given color appears infinitely often \cite{Harel,Alten}. As a
consequence, the satisfiability of MSO-formulas is at least $\Sigma_1^1$-hard.

In this paper, we will prove how various classes of formula correspond to
well known classes of subshifts. Some of the results of this paper were
already presented in \cite{JeTh}.
\section{Symbolic Spaces and Logic}
\label{sec:def}

\subsection{Configurations}
%Let ${d\geq 1}$ be a fixed integer and 
Consider the discrete lattice $\ZZ^2$. For any finite set $Q$, a $Q$-configuration is a function from
$\ZZ^2$ to $Q$. $Q$ may be seen as a set of \emph{colors} or \emph{states}.
An element of $\ZZ^2$ will be called a \emph{cell}. A configuration will
usually be denoted $C,M$ or $N$.

Fig.~\ref{conf:example} shows an example of two different
configurations of $\ZZ^2$ over a set $Q$ of $5$ colors. As a
configuration is infinite, only a finite fragment of the
configurations is represented in the figure. We choose not to
represent which cell of the picture is the origin $(0,0)$. This will
indeed be of no importance as we use only translation invariant
properties.

For any $z\in\ZZ^2$ we denote by $\sigma_z$ the \emph{shift} map of vector $z$,
\textit{i.e.} the function from $Q$-configurations to
$Q$-configurations such that for all ${C\in Q^{\ZZ^2}}$:
\[\forall z'\in\ZZ^2,\ \sigma_z(C)(z') = C(z'-z).\]

% lime olive white blue gray
\newcommand\clime{white!30!black}
\newcommand\colive{white!50!black}
\newcommand\cblue{black!90!white}
\newcommand\cdark{black!80!white}

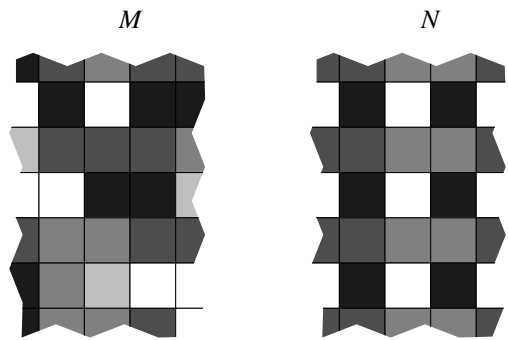
\begin{figure}[!htb]
\[
	\begin{array}{cc}
M & N\\
        \begin{tikzpicture}[scale=.6]
                \clip (0,1) rectangle (6,8);
                \clip[decorate, decoration={zigzag,segment length=9mm}] (0.5,7.5)   -- (4.5,7.5) -- (4.5,1.5)--(0.5,1.5) -- (0.5,7.5);              
                \filldraw[fill=\cblue]    (0,7) rectangle +(1,1);
                \filldraw[fill=\clime]  (1,7) rectangle +(1,1);
                \filldraw[fill=\colive]   (2,7) rectangle +(1,1);
                \filldraw[fill=\clime]   (3,7) rectangle +(1,1);
                \filldraw[fill=\clime]   (4,7) rectangle +(1,1);  
                \filldraw[fill=white] (0,6) rectangle +(1,1);
                \filldraw[fill=\cblue]    (1,6) rectangle +(1,1);
                \filldraw[fill=white] (2,6) rectangle +(1,1);
                \filldraw[fill=\cblue]    (3,6) rectangle +(1,1);
                \filldraw[fill=\cblue]    (4,6) rectangle +(1,1);
                \filldraw[fill=\clime]  (1,5) rectangle +(1,1);
                \filldraw[fill=\clime]   (2,5) rectangle +(1,1);
                \filldraw[fill=\clime]  (3,5) rectangle +(1,1);
                \filldraw[fill=\colive]   (4,5) rectangle +(1,1);
                \filldraw[fill=lightgray] (0,5) rectangle +(1,1);
                \filldraw[fill=white] (1,4) rectangle +(1,1);
                \filldraw[fill=\cblue]    (2,4) rectangle +(1,1);
                \filldraw[fill=\cblue]    (3,4) rectangle +(1,1);
                \filldraw[fill=white] (0,4) rectangle +(1,1);
                \filldraw[fill=lightgray]   (4,4) rectangle +(1,1);
                \filldraw[fill=\colive]   (1,3) rectangle +(1,1);
                \filldraw[fill=\colive]   (2,3) rectangle +(1,1);
                \filldraw[fill=\clime]  (3,3) rectangle +(1,1);
                \filldraw[fill=\clime]  (4,3) rectangle +(1,1);
                \filldraw[fill=\clime]  (0,3) rectangle +(1,1);           
                \filldraw[fill=\colive]   (1,2) rectangle +(1,1);
                \filldraw[fill=lightgray]   (2,2) rectangle +(1,1);
                \filldraw[fill=white]  (3,2) rectangle +(1,1);
                \filldraw[fill=\cblue]    (0,2) rectangle +(1,1);
                \filldraw[fill=white]    (4,2) rectangle +(1,1);
                \filldraw[fill=\colive]   (1,1) rectangle +(1,1);
                \filldraw[fill=\colive]   (2,1) rectangle +(1,1);
                \filldraw[fill=\clime]  (3,1) rectangle +(1,1);
                \filldraw[fill=white]  (4,1) rectangle +(1,1);
                \filldraw[fill=\cblue]    (0,1) rectangle +(1,1);
%\pgftext[at=\pgfpoint{2.5cm}{4.5cm}]{$a$}
        \end{tikzpicture}
&
        \begin{tikzpicture}[scale=.6]
                \clip (0,1) rectangle (6,8);
                \clip[decorate, decoration={zigzag,segment length=9mm}] (0.5,7.5)   -- (4.5,7.5) -- (4.5,1.5)--(0.5,1.5) -- (0.5,7.5);
                \filldraw[fill=\clime]    (0,7) rectangle +(1,1);
                \filldraw[fill=\clime]  (1,7) rectangle +(1,1);
                \filldraw[fill=\colive]   (2,7) rectangle +(1,1);
                \filldraw[fill=\colive]   (3,7) rectangle +(1,1);
                \filldraw[fill=\clime]  (4,7) rectangle +(1,1);   
                \filldraw[fill=white] (0,6) rectangle +(1,1);
                \filldraw[fill=\cblue]    (1,6) rectangle +(1,1);
                \filldraw[fill=white] (2,6) rectangle +(1,1);
                \filldraw[fill=\cblue]    (3,6) rectangle +(1,1);
                \filldraw[fill=white]    (4,6) rectangle +(1,1);
                \filldraw[fill=\clime]    (0,5) rectangle +(1,1);
                \filldraw[fill=\clime]  (1,5) rectangle +(1,1);
                \filldraw[fill=\colive]   (2,5) rectangle +(1,1);
                \filldraw[fill=\colive]   (3,5) rectangle +(1,1);
                \filldraw[fill=\clime]  (4,5) rectangle +(1,1);   
                \filldraw[fill=white] (0,4) rectangle +(1,1);
                \filldraw[fill=\cblue]    (1,4) rectangle +(1,1);
                \filldraw[fill=white] (2,4) rectangle +(1,1);
                \filldraw[fill=\cblue]    (3,4) rectangle +(1,1);
                \filldraw[fill=white]    (4,4) rectangle +(1,1);
                \filldraw[fill=\clime]    (0,3) rectangle +(1,1);
                \filldraw[fill=\clime]  (1,3) rectangle +(1,1);
                \filldraw[fill=\colive]   (2,3) rectangle +(1,1);
                \filldraw[fill=\colive]   (3,3) rectangle +(1,1);
                \filldraw[fill=\clime]  (4,3) rectangle +(1,1);   
                \filldraw[fill=white] (0,2) rectangle +(1,1);
                \filldraw[fill=\cblue]    (1,2) rectangle +(1,1);
                \filldraw[fill=white] (2,2) rectangle +(1,1);
                \filldraw[fill=\cblue]    (3,2) rectangle +(1,1);
                \filldraw[fill=white]    (4,2) rectangle +(1,1);
                \filldraw[fill=\clime]    (0,1) rectangle +(1,1);
                \filldraw[fill=\clime]  (1,1) rectangle +(1,1);
                \filldraw[fill=\colive]   (2,1) rectangle +(1,1);
                \filldraw[fill=\colive]   (3,1) rectangle +(1,1);
                \filldraw[fill=\clime]  (4,1) rectangle +(1,1);
%\pgftext[at=\pgfpoint{1.5cm}{3.5cm}]{$a$}
        \end{tikzpicture}
\end{array}
\]
\caption{Two configurations}
\label{conf:example}
\end{figure}

\begin{figure}[!htb]
\[
        \begin{tikzpicture}[scale=.6]
                \filldraw[fill=\clime]   (3,7) rectangle +(1,1);
                \filldraw[fill=\clime]   (4,7) rectangle +(1,1);  
                \filldraw[fill=white] (2,6) rectangle +(1,1);
                \filldraw[fill=\cblue]    (3,6) rectangle +(1,1);
        \end{tikzpicture}
\]
\caption{A pattern $P$. $P$ appears in $M$ but presumably not in $N$}
\label{pattern:example}
\end{figure}
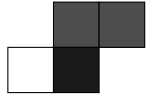

A \emph{pattern} is a partial configuration. A pattern ${P :
  X\rightarrow Q}$ where $X \subseteq \ZZ^2$ occurs in ${C\in Q^{\ZZ^2}}$ at position $z_0$ if
\[\forall z\in X,\ C(z_0+z)=P(z).\]
We say that $P$ occurs in $C$ if it occurs at some position in $C$. As
an example the pattern $P$ of Fig~\ref{pattern:example} occurs in the
configuration $M$ but not in $N$ (or more accurately not on the finite
fragment of $N$ depicted in the figure).  A finite pattern is a
partial configuration of finite domain. All patterns in the following
will be finite. The \emph{language} $\lang(C)$ of a configuration $C$
is the set of finite patterns that occur in $C$. We naturally extend
this notion to sets of configurations.

A \emph{subshift} is a natural concept that captures both the notion of
\emph{uniformity} and \emph{locality}: the only description
``available'' from a configuration $C$ is the finite patterns it
contains, that is $\lang(C)$.  Given a set $\cal F$ of patterns, let
$X_{\cal F}$ be the set of all configurations where no patterns of
$\cal F$ occurs.
\[
	X_{\cal F} = \{ C | \lang(C)\cap \mathcal{F}=\emptyset \}
\]
$\cal F$ is usually called the set of forbidden patterns or the
\emph{forbidden language}. A set of the form $X_{\cal F}$ is called a
\emph{subshift}.

 A subshift can be equivalentely defined by topology
  considerations. Endow the set of configurations $Q^{\ZZ^2}$ with the
  product topology: A sequence $(C_n)_{n \in \NN}$ of configurations
  converges to a configuration $C$ if the sequence ultimately agree
  with $C$ on every $z \in \ZZ^2$. Then a subshift is a closed subset
  of $Q^{\ZZ^2}$ also closed by shift maps. 
%\newpage
\begin{figure}
	\begin{center}
        \begin{tikzpicture}[scale=0.7]
			\draw (0,10) node {Patterns:};
                \filldraw[fill=\cdark]    (2,10) rectangle +(1,1);
                \filldraw[fill=gray]  (3,10) rectangle +(1,1);

                \filldraw[fill=gray]    (7,11) rectangle +(1,1);
                \filldraw[fill=\cdark]  (7,10) rectangle +(1,1);

                \filldraw[fill=gray]  (12,10) rectangle +(1,1);
                \filldraw[fill=\cdark]  (12,11) rectangle +(1,1);
                \filldraw[fill=\cdark]  (13,10) rectangle +(1,1);

	% Quart rouge
       \clip (0,.5) rectangle (16,8);
 			\draw (6,2.5) node {$B$};
 			\draw (8.5,2.5) node {$C$};
 			\draw (11,2.5) node {$D$};
 			\draw (13.5,2.5) node {$E$};

	\begin{scope}	   
 			\draw (2.5,1) node {$A$};
		
                \clip[decorate, decoration={zigzag,segment length=4mm}]  (0.5,7.5)   -- (4.5,7.5) -- (4.5,1.5)--(0.5,1.5) -- (0.5,7.5);              
{               \filldraw[fill=gray]    (0,7) rectangle +(1,1);
                \filldraw[fill=gray]  (1,7) rectangle +(1,1);}               
                \filldraw[fill=\cdark]   (2,7) rectangle +(1,1);
                \filldraw[fill=\cdark]   (3,7) rectangle +(1,1);
                \filldraw[fill=\cdark]   (4,7) rectangle +(1,1);
                \filldraw[fill=\cdark] (2,6) rectangle +(1,1);
                \filldraw[fill=\cdark]    (3,6) rectangle +(1,1);
                \filldraw[fill=\cdark]    (4,6) rectangle +(1,1);          
                \filldraw[fill=\cdark]    (2,4) rectangle +(1,1);
                \filldraw[fill=\cdark]    (3,4) rectangle +(1,1);
                \filldraw[fill=\cdark]   (4,4) rectangle +(1,1);           
                \filldraw[fill=\cdark]   (2,5) rectangle +(1,1);
                \filldraw[fill=\cdark]  (3,5) rectangle +(1,1);
                \filldraw[fill=\cdark]   (4,5) rectangle +(1,1);           
          
{
                \filldraw[fill=gray] (0,6) rectangle +(1,1);
                \filldraw[fill=gray]    (1,6) rectangle +(1,1);}
{               
                \filldraw[fill=gray] (0,5) rectangle +(1,1);
                \filldraw[fill=gray]  (1,5) rectangle +(1,1);}
{               
                \filldraw[fill=gray] (0,4) rectangle +(1,1);
                \filldraw[fill=gray] (1,4) rectangle +(1,1);}
{
                \filldraw[fill=gray]  (0,3) rectangle +(1,1);                
                \filldraw[fill=gray]   (1,3) rectangle +(1,1);
                }
{               \filldraw[fill=gray]   (2,3) rectangle +(1,1);
                \filldraw[fill=gray]  (3,3) rectangle +(1,1);
                \filldraw[fill=gray]  (4,3) rectangle +(1,1);}
{         
                \filldraw[fill=gray]    (0,2) rectangle +(1,1);
                \filldraw[fill=gray]   (1,2) rectangle +(1,1);}
{               \filldraw[fill=gray]   (2,2) rectangle +(1,1);
                \filldraw[fill=gray]  (3,2) rectangle +(1,1);
                \filldraw[fill=gray]    (4,2) rectangle +(1,1);}
{               \filldraw[fill=gray]    (0,1) rectangle +(1,1);
                \filldraw[fill=gray]   (1,1) rectangle +(1,1);
                }
{               \filldraw[fill=gray]   (2,1) rectangle +(1,1);
                \filldraw[fill=gray]  (3,1) rectangle +(1,1);
                \filldraw[fill=gray]  (4,1) rectangle +(1,1);}
\end{scope}			  
\begin{scope}[scale=0.5,xshift=10cm, yshift=5cm]
	% Demi rouge
       \clip (0,1) rectangle (6,8);
                \clip[decorate, decoration={zigzag,segment length=4mm}]  (0.5,7.5)   -- (4.5,7.5) -- (4.5,1.5)--(0.5,1.5) -- (0.5,7.5);              
{               \filldraw[fill=gray]    (0,7) rectangle +(1,1);
                \filldraw[fill=gray]  (1,7) rectangle +(1,1);}               
                \filldraw[fill=\cdark]   (2,7) rectangle +(1,1);
                \filldraw[fill=\cdark]   (3,7) rectangle +(1,1);
                \filldraw[fill=\cdark]   (4,7) rectangle +(1,1);
                \filldraw[fill=\cdark] (2,6) rectangle +(1,1);
                \filldraw[fill=\cdark]    (3,6) rectangle +(1,1);
                \filldraw[fill=\cdark]    (4,6) rectangle +(1,1);          
                \filldraw[fill=\cdark]    (2,4) rectangle +(1,1);
                \filldraw[fill=\cdark]    (3,4) rectangle +(1,1);
                \filldraw[fill=\cdark]   (4,4) rectangle +(1,1);           
                \filldraw[fill=\cdark]   (2,5) rectangle +(1,1);
                \filldraw[fill=\cdark]  (3,5) rectangle +(1,1);
                \filldraw[fill=\cdark]   (4,5) rectangle +(1,1);           
          
{
                \filldraw[fill=gray] (0,6) rectangle +(1,1);
                \filldraw[fill=gray]    (1,6) rectangle +(1,1);}
{               
                \filldraw[fill=gray] (0,5) rectangle +(1,1);
                \filldraw[fill=gray]  (1,5) rectangle +(1,1);}
{               
                \filldraw[fill=gray] (0,4) rectangle +(1,1);
                \filldraw[fill=gray] (1,4) rectangle +(1,1);}
{
                \filldraw[fill=gray]  (0,3) rectangle +(1,1);                
                \filldraw[fill=gray]   (1,3) rectangle +(1,1);
                }
{               \filldraw[fill=\cdark]   (2,3) rectangle +(1,1);
                \filldraw[fill=\cdark]  (3,3) rectangle +(1,1);
                \filldraw[fill=\cdark]  (4,3) rectangle +(1,1);}
{         
                \filldraw[fill=gray]    (0,2) rectangle +(1,1);
                \filldraw[fill=gray]   (1,2) rectangle +(1,1);}
{               \filldraw[fill=\cdark]   (2,2) rectangle +(1,1);
                \filldraw[fill=\cdark]  (3,2) rectangle +(1,1);
                \filldraw[fill=\cdark]    (4,2) rectangle +(1,1);}
{               \filldraw[fill=gray]    (0,1) rectangle +(1,1);
                \filldraw[fill=gray]   (1,1) rectangle +(1,1);
                }
{               \filldraw[fill=\cdark]   (2,1) rectangle +(1,1);
                \filldraw[fill=\cdark]  (3,1) rectangle +(1,1);
                \filldraw[fill=\cdark]  (4,1) rectangle +(1,1);}
\end{scope}			  
\begin{scope}[scale=0.5,xshift=15cm, yshift=5cm]
	% Demi rouge
       \clip (0,-10) rectangle (6,8);
                \clip[decorate, decoration={zigzag,segment length=4mm}]  (0.5,7.5)   -- (4.5,7.5) -- (4.5,1.5)--(0.5,1.5) -- (0.5,7.5);              
                \filldraw[fill=\cdark]   (2,7) rectangle +(1,1);

                \filldraw[fill=\cdark]   (3,7) rectangle +(1,1);
                \filldraw[fill=\cdark]   (4,7) rectangle +(1,1);
                \filldraw[fill=\cdark] (2,6) rectangle +(1,1);
                \filldraw[fill=\cdark]    (3,6) rectangle +(1,1);
                \filldraw[fill=\cdark]    (4,6) rectangle +(1,1);          
                \filldraw[fill=\cdark]    (2,4) rectangle +(1,1);
                \filldraw[fill=\cdark]    (3,4) rectangle +(1,1);
                \filldraw[fill=\cdark]   (4,4) rectangle +(1,1);           
                \filldraw[fill=\cdark]   (2,5) rectangle +(1,1);
                \filldraw[fill=\cdark]  (3,5) rectangle +(1,1);
                \filldraw[fill=\cdark]   (4,5) rectangle +(1,1);           

{               \filldraw[fill=\cdark]    (0,7) rectangle +(1,1);
                \filldraw[fill=\cdark]  (1,7) rectangle +(1,1);}                         
{
                \filldraw[fill=\cdark] (0,6) rectangle +(1,1);
                \filldraw[fill=\cdark]    (1,6) rectangle +(1,1);}
{               
                \filldraw[fill=\cdark] (0,5) rectangle +(1,1);
                \filldraw[fill=\cdark]  (1,5) rectangle +(1,1);}
{               
                \filldraw[fill=\cdark] (0,4) rectangle +(1,1);
                \filldraw[fill=\cdark] (1,4) rectangle +(1,1);}
{
                \filldraw[fill=gray]  (0,3) rectangle +(1,1);                
                \filldraw[fill=gray]   (1,3) rectangle +(1,1);
                }
{               \filldraw[fill=gray]   (2,3) rectangle +(1,1);
                \filldraw[fill=gray]  (3,3) rectangle +(1,1);
                \filldraw[fill=gray]  (4,3) rectangle +(1,1);}
{         
                \filldraw[fill=gray]    (0,2) rectangle +(1,1);
                \filldraw[fill=gray]   (1,2) rectangle +(1,1);}
{               \filldraw[fill=gray]   (2,2) rectangle +(1,1);
                \filldraw[fill=gray]  (3,2) rectangle +(1,1);
                \filldraw[fill=gray]    (4,2) rectangle +(1,1);}
{               \filldraw[fill=gray]    (0,1) rectangle +(1,1);
                \filldraw[fill=gray]   (1,1) rectangle +(1,1);
                }
{               \filldraw[fill=gray]   (2,1) rectangle +(1,1);
                \filldraw[fill=gray]  (3,1) rectangle +(1,1);
                \filldraw[fill=gray]  (4,1) rectangle +(1,1);}
\end{scope}
\begin{scope}[scale=0.5,xshift=20cm, yshift=5cm]
	% Tout rouge
       \clip (0,0) rectangle (6,8);
                \clip[decorate, decoration={zigzag,segment length=4mm}]  (0.5,7.5)   -- (4.5,7.5) -- (4.5,1.5)--(0.5,1.5) -- (0.5,7.5);              
                \filldraw[fill=\cdark]   (2,7) rectangle +(1,1);
                \filldraw[fill=\cdark]   (3,7) rectangle +(1,1);
                \filldraw[fill=\cdark]   (4,7) rectangle +(1,1);
                \filldraw[fill=\cdark] (2,6) rectangle +(1,1);
                \filldraw[fill=\cdark]    (3,6) rectangle +(1,1);
                \filldraw[fill=\cdark]    (4,6) rectangle +(1,1);          
                \filldraw[fill=\cdark]    (2,4) rectangle +(1,1);
                \filldraw[fill=\cdark]    (3,4) rectangle +(1,1);
                \filldraw[fill=\cdark]   (4,4) rectangle +(1,1);           
                \filldraw[fill=\cdark]   (2,5) rectangle +(1,1);
                \filldraw[fill=\cdark]  (3,5) rectangle +(1,1);
                \filldraw[fill=\cdark]   (4,5) rectangle +(1,1);           

{               \filldraw[fill=\cdark]    (0,7) rectangle +(1,1);
                \filldraw[fill=\cdark]  (1,7) rectangle +(1,1);}                         
{
                \filldraw[fill=\cdark] (0,6) rectangle +(1,1);
                \filldraw[fill=\cdark]    (1,6) rectangle +(1,1);}
{               
                \filldraw[fill=\cdark] (0,5) rectangle +(1,1);
                \filldraw[fill=\cdark]  (1,5) rectangle +(1,1);}
{               
                \filldraw[fill=\cdark] (0,4) rectangle +(1,1);
                \filldraw[fill=\cdark] (1,4) rectangle +(1,1);}
{
                \filldraw[fill=\cdark]  (0,3) rectangle +(1,1);                
                \filldraw[fill=\cdark]   (1,3) rectangle +(1,1);
                }
{               \filldraw[fill=\cdark]   (2,3) rectangle +(1,1);
                \filldraw[fill=\cdark]  (3,3) rectangle +(1,1);
                \filldraw[fill=\cdark]  (4,3) rectangle +(1,1);}
{         
                \filldraw[fill=\cdark]    (0,2) rectangle +(1,1);
                \filldraw[fill=\cdark]   (1,2) rectangle +(1,1);}
{               \filldraw[fill=\cdark]   (2,2) rectangle +(1,1);
                \filldraw[fill=\cdark]  (3,2) rectangle +(1,1);
                \filldraw[fill=\cdark]    (4,2) rectangle +(1,1);}
{               \filldraw[fill=\cdark]    (0,1) rectangle +(1,1);
                \filldraw[fill=\cdark]   (1,1) rectangle +(1,1);
                }
{               \filldraw[fill=\cdark]   (2,1) rectangle +(1,1);
                \filldraw[fill=\cdark]  (3,1) rectangle +(1,1);
                \filldraw[fill=\cdark]  (4,1) rectangle +(1,1);}
\end{scope}
\begin{scope}[scale=0.5,xshift=25cm, yshift=5cm]
	% Tout bleu
       \clip (0,1) rectangle (6,8);
                \clip[decorate, decoration={zigzag,segment length=4mm}]  (0.5,7.5)   -- (4.5,7.5) -- (4.5,1.5)--(0.5,1.5) -- (0.5,7.5);              
                \filldraw[fill=gray]   (2,7) rectangle +(1,1);
                \filldraw[fill=gray]   (3,7) rectangle +(1,1);
                \filldraw[fill=gray]   (4,7) rectangle +(1,1);
                \filldraw[fill=gray] (2,6) rectangle +(1,1);
                \filldraw[fill=gray]    (3,6) rectangle +(1,1);
                \filldraw[fill=gray]    (4,6) rectangle +(1,1);          
                \filldraw[fill=gray]    (2,4) rectangle +(1,1);
                \filldraw[fill=gray]    (3,4) rectangle +(1,1);
                \filldraw[fill=gray]   (4,4) rectangle +(1,1);           
                \filldraw[fill=gray]   (2,5) rectangle +(1,1);
                \filldraw[fill=gray]  (3,5) rectangle +(1,1);
                \filldraw[fill=gray]   (4,5) rectangle +(1,1);           

{               \filldraw[fill=gray]    (0,7) rectangle +(1,1);
                \filldraw[fill=gray]  (1,7) rectangle +(1,1);}                         
{
                \filldraw[fill=gray] (0,6) rectangle +(1,1);
                \filldraw[fill=gray]    (1,6) rectangle +(1,1);}
{               
                \filldraw[fill=gray] (0,5) rectangle +(1,1);
                \filldraw[fill=gray]  (1,5) rectangle +(1,1);}
{               
                \filldraw[fill=gray] (0,4) rectangle +(1,1);
                \filldraw[fill=gray] (1,4) rectangle +(1,1);}
{
                \filldraw[fill=gray]  (0,3) rectangle +(1,1);                
                \filldraw[fill=gray]   (1,3) rectangle +(1,1);
                }
{               \filldraw[fill=gray]   (2,3) rectangle +(1,1);
                \filldraw[fill=gray]  (3,3) rectangle +(1,1);
                \filldraw[fill=gray]  (4,3) rectangle +(1,1);}
{         
                \filldraw[fill=gray]    (0,2) rectangle +(1,1);
                \filldraw[fill=gray]   (1,2) rectangle +(1,1);}
{               \filldraw[fill=gray]   (2,2) rectangle +(1,1);
                \filldraw[fill=gray]  (3,2) rectangle +(1,1);
                \filldraw[fill=gray]    (4,2) rectangle +(1,1);}
{               \filldraw[fill=gray]    (0,1) rectangle +(1,1);
                \filldraw[fill=gray]   (1,1) rectangle +(1,1);
                }
{               \filldraw[fill=gray]   (2,1) rectangle +(1,1);
                \filldraw[fill=gray]  (3,1) rectangle +(1,1);
                \filldraw[fill=gray]  (4,1) rectangle +(1,1);}
\end{scope}			  
\end{tikzpicture}
\end{center}
\caption{\label{fig:sft}A (finite) set of forbidden patterns $\cal F$ and the tilings it
  generates}
\end{figure}

\renewcommand\pred[2]{P_{\textbf{#1}} (#2)}
\renewcommand\tcolor[1]{\textbf{#1}}
\renewcommand\clime{A}
\renewcommand\colive{B}
\renewcommand\cblue{C}
\renewcommand\cdark{D}

\begin{example}
	\label{ex:corner}
	Consider the three forbidden patterns of figure \ref{fig:sft}
        and denote by \textbf{D} the dark color and \textbf{L} the
        light color. The first one says that we cannot find a
        \tcolor{\cdark} point at the left of a \tcolor{L} point. This
        can be interpreted as follows: every time we find a
        \tcolor{\cdark} point, then all the points at the right of it
        are also \tcolor{\cdark}. With the second forbidden pattern,
        we deduce that every time we find a \tcolor{\cdark} point,
        then the entire quarter of plane on the above right of it is
        also filled with \tcolor{\cdark} points.  The third pattern
        ensures us that every configuration contains at most one
        quarter of plane of color \tcolor{\cdark} : if it contains two
        such quarters of plane, then there must be a bigger quarter of
        plane that contains both.

Hence a typical configuration looks like $A$.
Other possible configurations are $B,C,D,E$. They
correspond to extremal situations where the corner of the quarter of plane
is situated respectively at $(0, -\infty)$, $(-\infty, 0)$, $(-\infty,
-\infty)$ et $(+\infty, +\infty)$
\end{example}
\begin{example}
	\label{ex:one}
Consider the set of colors $\left\{\tcolor{\cdark}, \tcolor{W}\right\}$ and $\cal F$ to
be the set of patterns that contains two $\tcolor{\cdark}$ points or more.

Then $X_{\cal F}$ contains configurations with at most one $\tcolor{\cdark}$ point. Up
to shift, $X_{\cal F}$ contains then two configurations: the all
$\tcolor{W}$-one, and one where only one point is $\tcolor{\cdark}$
and all others are $\tcolor{W}$.

\end{example}	
\newpage
A \emph{subshift of finite type} (or \emph{tiling}) \correc{is a subshift that can be defined via a finite
set $\cal F$ of forbidden patterns}: it is the set of configurations $C$ such that no pattern
in $\cal F$ occurs in $C$. If all patterns of $\cal F$ \correc{fit in a $n\times n$ square},
this means that we only have to see a configuration through a window of
size $n\times n$ to know if it is a tiling, hence the locality.
Example~\ref{ex:corner} is a subshift of finite type. It can be proven that
Example~\ref{ex:one} is not.

Given two state sets $Q_1$ and $Q_2$, a projection is a map ${\pi:
  Q_1\rightarrow Q_2}$. We naturally extend it to ${\pi :
  Q_1^{\ZZ^2}\rightarrow Q_2^{\ZZ^2}}$ by $\pi(C)(z) = \pi(C(z)).$ A
\emph{sofic subshift} of state set $Q_2$ is the image by some
projection $\pi$ of some subshift of finite type of state set
$Q_1$. It is also a subshift {\cR(clearly closed by shift maps, and
  topologically closed because projections are continuous maps on a
  compact space)}. A sofic subshift is a natural object in tiling
theory, although quite never mentioned explicitly. It represents the
concept of \emph{decoration}: some of the tiles we assemble to obtain
the tilings may be decorated, but we forgot the decoration when we
observe the tiling.%  \newpage
\begin{example}
Consider the following variant of Example~\ref{ex:corner}: tilings are
exactly the same except that the corner of the quarter of plane in $A$ is of
a different color \tcolor{W}. It is easy to see that this variant
defines a subshift of finite type $X$ (with a few more forbidden patterns).

Now consider the following map:
\[
	\pi: 
	\begin{array}{rrr} 
	\tcolor{L} &\mapsto& \tcolor{W}\\
	\tcolor{\cdark} &\mapsto& \tcolor{W}\\
	\tcolor{W} &\mapsto& \tcolor{\cdark}\\		
	\end{array}	
	\]
Then $B,C,D,E$ will become under $\pi$ of color \tcolor{W}, while $A$
will become a configuration with exactly one \tcolor{\cdark}, all other
points being \tcolor{W}.

As a consequence, $\pi(X)$ is exactly Example~\ref{ex:one}.
Example~\ref{ex:one} is thus a sofic subshift.

\end{example}
\subsection{Structures}
\label{sec:struc}

%{\cR From now on, we restrict to dimension 2.}  %\TODO[on y est deja avant]

A configuration will
be seen in this article as an infinite structure.  The signature
$\tau$ contains four unary maps $\north$, $\south$, $\east$, $\west$
and a predicate $P_c$ for each color $c \in Q$.

A configuration $M$ will be seen as a structure $\mathfrak{M}$ in the
following way: 
\begin{itemize}
	\item The elements of $\mathfrak{M}$ are the points of $\ZZ^2$.
	\item $\north$ is interpreted by $\north^\mathfrak{M}((x,y)) = (x,y+1)$, $\east$ is
	  interpreted by
	  ${\east^\mathfrak{M}((x,y)) = (x+1,y)}$. $\south^\mathfrak{M}$ and
	  $\west^\mathfrak{M}$ are interpreted similarly
	\item $P^\mathfrak{M}_c((x,y))$ is true if and only if the point at
	  coordinate $(x,y)$ is of color $c$, that is if $M(x,y) = c$.
\end{itemize}

As an example, the configuration $M$ of Fig.~\ref{conf:example} has
three consecutive cells with the color $\tcolor{\clime}$ (second row
from the top, colors are denoted \textbf{A}, \textbf{B}, \textbf{C}, \textbf{D}, \textbf{E} below). That is,
the following formula is true:
\[
	\mathfrak{M} \models \exists z, \pred{\clime}{z} \wedge \pred{\clime}{\east(z)}
	\wedge \pred{\clime}{\east(\east(z))}
\]

As another example, the following formula states that the configuration has
a vertical period of $2$ (the color in the cell $(x,y)$ is the same as the
color in the cell $(x,y+2)$). The formula is false in the structure $\mathfrak{M}$ and
true in the structure $\mathfrak{N}$ (if the reader chose to color the
cells of $N$ not shown in the picture correctly):

\[	
	\forall z,
\left\{
\begin{array}{c}
	\pred{\clime}{z} \implies \pred{\clime}{\north(\north(z))}\\
	\pred{\colive}{z} \implies \pred{\colive}{\north(\north(z))}\\
	\pred{\cblue}{z} \implies \pred{\cblue}{\north(\north(z))}\\
	\pred{D}{z} \implies \pred{D}{\north(\north(z))}\\
	\pred{E}{z} \implies \pred{E}{\north(\north(z))}\\
\end{array}
\right.
	\]

\correc{\textit{Remark.} The choice of unary function (north, south,
east, west) instead of binary relations in the signature above is
important because it allows a simple characterization of important
classes of subshifts (see theorem~\ref{sft} below). % \correct{
  This particular
theorem %\xout{Such a characterization}}
would fail with binary relations in the signature
instead of unary functions. % \correct{
  Other theorems would be still valid.}%}

\subsection{Monadic Second-Order Logic}
\label{sec:mso}
This paper studies connection between subshifts (seen as structures
as explained above) and monadic second order sentences. First order
variables ($x$, $y$, $z$, ...) are interpreted as points of $\ZZ^2$
and (monadic) second order variables ($X$, $Y$, $Z$, ...) as subsets
of~$\ZZ^2$.

Monadic second order formulas \correc{($\phi$, $\psi$, ...)} are defined as follows:
\begin{itemize}
\item a term is either a first-order variable or a function ($\south$,
  $\north$, $\east$, $\west$) applied to a term ;
\item atomic formulas are of the form $t_1=t_2$
  or $X(t_1)$ where $t_1$ and $t_2$ are terms and $X$ is either a second order variable or a color predicate ;
\item formulas are build up from atomic formulas by means of boolean
  connectives and quantifiers $\exists$ and $\forall$ (which can be
  applied either to first-order variables or second order variables).
\end{itemize}

A formula is \emph{closed} if no variable occurs free in it. A formula
is FO if no second-order quantifier occurs in it. A formula is EMSO if
it is of the form 
\[\exists X_1,\ldots,\exists X_n, \phi(X)\] where $\phi$ is FO.  Given a
formula $\phi(X_1,\ldots,X_n)$ with no free first-order variable and
having only $X_1,\ldots, X_n$ as free second-order variables, a
configuration $M$ together with subsets ${E_1,\ldots, E_n}$ is a model
of $\phi(X_1,\ldots,X_n)$, denoted
\[(M, E_1,\ldots, E_n) \models \phi(X_1,\ldots,X_n),\] if $\phi$ is
satisfied (in the usual sense) when $M$ is interpreted as $\mathfrak
M$ (see previous section) and $E_i$ interprets $X_i$.

%\newpage
\subsection{Definability}
This paper studies the following problems: Given a formula $\phi$ of
some logic, what can be said of the configurations that satisfy
$\phi$? Conversely, given a subshift, what kind of formula can
characterise it?
\begin{defn}
	A set $S$ of $Q$-configurations is defined by $\phi$ if
	\[ S = \left\{ M \in Q^{\ZZ^2} \middle| \mathfrak{M} \models \phi \right\}\]	
	Two formulas $\phi$ and $\phi'$ are equivalent iff they define the same
	set of configurations.
	
    A set $S$ is $\cal C$-definable if it is defined by a formula  $\phi \in \cal C$.		
\end{defn}
It is easy to see that Example~\ref{ex:corner} is defined by the formula 
\newcommand\place{\vrule height 12pt depth 12pt width 0pt}
\[
	\phi : 
	\left\{
		  \begin{array}{l}
	\forall x, \place \neg\left(\pred{\cdark}{x} \wedge
	\pred{L}{\east(x)}\right)\\
	\forall x, \place \neg\left(\pred{\cdark}{x} \wedge
	\pred{L}{\north(x)}\right)\\
	\forall x, \place \neg\left(\pred{L}{x} \wedge \pred{\cdark}{\east(x)} \wedge
	\pred{L}{\north(x)}\right)\\
	\end{array}
	\right.
	\]
or equivalently by the formula
\[
	\phi' :
	\forall x, \pred{\cdark}{x} \iff \left(\pred{\cdark}{\east(x)} \wedge
	\pred{\cdark}{\north(x)}\right)	
	\]
We will see some variants of formula $\phi'$ appear in a few theorems below.

Example~\ref{ex:one} is defined by the formula
\[
	\psi: \forall x,y,  \left(\pred{\cdark}{x} \wedge \pred{\cdark}{y}\right) \implies x = y
	\]
Note that a definable set is always closed by shift (a shift between $2$ configurations 
induces an isomorphism between corresponding structures).
It is not always closed: The set of $\{\tcolor{\clime}, \tcolor{E}\}$-configurations 
  defined by the formula $\phi: {\exists z, \pred{\clime}{z}}$ contains all
  configurations except the all-white one, hence is not closed.

  When we are dealing with MSO formulas, the following remark is
  useful: second-order quantifiers may be represented as projection
  operations on sets of configurations. We formalize now this notion.

If $\pi : Q_1 \mapsto Q_2$ is a projection and $S$ is a set of
$Q_1$-configurations, we define the two following operators:
\[
	\begin{array}{rcl}
	E(\pi)(S) &=&\displaystyle \left\{ M \in (Q_2)^{\ZZ^2} \middle|  \exists N \in	  (Q_1)^{\ZZ^2}, \pi(N) = M \wedge N \in S \right\}\\[2ex]
	A(\pi)(S) &=&\displaystyle \left\{ M \in (Q_2)^{\ZZ^2} \middle|  \forall N \in	  (Q_1)^{\ZZ^2}, \pi(N) = M \implies N \in S\right\}
	\end{array}
	\]

Note that $A$ is a dual of $E$, that is $A(\pi)(S) = {^cE}(\pi)({^cS})$ where
$^c$ represents complementation.

\begin{prop}
	\mbox{}
\label{project}	
\begin{itemize}
	\item
	A set $S$ of $Q$-configurations is EMSO-definable if and only if there
	exists a set $S'$ of $Q'$ configurations and a map $\pi: Q' \mapsto Q$
	such that $S = E(\pi)(S')$ and $S'$ is FO-definable.
	\item
	  The class of MSO-definable sets is the closure of the class of
       FO-definable sets by the operators $E$ and $A$.
	   \end{itemize}
\end{prop}	
\begin{pf}[Sketch] \correc{Second item is a straightforward reformulation of the prenex normal form of MSO using operators $E$ and $A$.} We prove here only the first item.
\begin{itemize}
	\item Let $\phi = \exists X, \psi$ be a EMSO formula that defines a set $S$ of $Q$-configurations.
	  Let $Q' = Q \times \{ 0,1\}$ and $\pi$ be the canonical projection from $Q'$ to $Q$.

	  Consider the formula $\psi'$ obtained from $\psi$ by replacing $X(t)$
	  by $\vee_{c \in Q} P_{(c,1)}(t)$ and $P_c(t)$ by $P_{(c,0)}(t) \vee P_{(c,1)}(t)$.
	  
      Let $S'$ be a set of $Q'$ configurations defined by $\psi'$. Then is it
	  clear that $S = E(\pi)(S')$. The generalization to more than one
	  existential quantifier is straightforward.
	  
	\item Let $S = E(\pi)(S')$ be a set of $Q$ configurations, and $S'$
	  FO-definable by the formula $\phi$.	  	  
	  Denote by $c_1 \dots c_n$ the elements of $Q'$.
	  Consider the formula $\phi'$ obtained from $\phi$ where each $P_{c_i}$ is 
	  replaced by $X_i$.
      Let
	  
	  \[
\psi = \exists X_1, \dots, \exists X_n,
\left\{
	  \begin{array}{l}
   \forall z, \vee_i X_i(z)  \\
   \forall z, \wedge_{i\not=j} (\neg X_i(z) \vee \neg X_j(z)) \\
   \forall z, \wedge_i \left(X_i{z} \implies P_{\pi(c_i)} (z) \right)\\
   \phi'
   \end{array}\right.
  \]
  
  Then $\psi$ defines $S$.  Note that the formula $\psi$ constructed
  above is of the form ${\exists X_1, \dots, \exists X_n (\forall z, \psi'(z))
    \wedge \phi'}$.  This will be important later.\qed
\end{itemize}
\end{pf}

Second-order quantifications will then be regarded in this paper either as
projections operators or sets quantifiers.
%\newpage
\section{Hanf Locality Lemma and EMSO}

\label{sec:collapse}

The first-order logic has a property that makes it suitable to deal with
tilings and configurations: it is local. This is illustrated by Hanf's
lemma \cite{hanf65,EF:finmt,Libkin}.
A square pattern of radius $n$ is a pattern of domain $[-n,n] \times [n,n]$. 

\begin{defn}
	\label{hanf:def}
  Two $Q$-configurations $M$ and $N$ are $(n,k)$-equivalent if for each
  $Q$-square pattern $P$ of radius $n$:
  \begin{itemize}
	  \item If $P$ appears in $M$ \correc{at most} $k$ times, then $P$ appears the
		exact same number of times in $M$ and in $N$
	  \item If $P$ appears in $M$ more than $k$ times, then $P$ appears in
		$N$ more than $k$ times
  \end{itemize}	
\end{defn}
This notion is indeed an equivalence relation. Given $n$ and $k$, it
is clear that there is only finitely many equivalence classes for this
relation.
%\TODO[note que c'est vrai avec cette definition, mais
%ce n'etait pas vrai avec la def precedente]

\correc{Contrary to definition~\ref{hanf:def} above, Hanf's original formalism doesn't use square shapes (balls for the $\|\cdot\|_\infty$ norm) but lozenges (balls for the $\|\cdot\|_1$ norm). It makes essentially no difference and } the Hanf's local lemma can be reformulated in our context as follows \correc{(proofs using formalism of definition~\ref{hanf:def} appear in \cite{GiamRest})}. 
\begin{thm}
	\label{hanf:thm}
	For every FO formula $\phi$, there exists $(n,k)$ such that
	\begin{center}
	if $M$ and $N$ are $(n,k)$ equivalent, then
	$\mathfrak{M} \models \phi \iff \mathfrak{N} \models \phi$
	\end{center}
\end{thm}	

\begin{cor}
	Every FO-definable set is a (finite) union of some $(n,k)$-equivalence classes.
\end{cor}	
This is theorem 3.3 in  \cite{GiamRest}, stated for finite configurations.
Lemma 3.5 in the same paper gives a proof of Hanf's Local Lemma in our context.

Given $(P, k)$
we consider the set $S_{=k}(P)$ of all configurations such that the pattern
$P$ occurs exactly $k$ times ($k$ may be taken equal to $0$). The set $S_{\geq k}(P)$ is the set of all
configurations such that the pattern $P$ occurs $k$ times or more.

We may rephrase the preceding corollary as:
\begin{cor}
  \label{fosets}
	Every FO-definable set is a positive combination (i.e. unions and intersections) of some $S_{=k}(P)$ and some $S_{\geq k}(P)$
\end{cor}	

\begin{thm}
\label{thm:emso}
Every EMSO-definable set can be defined by a formula $\phi$ of the form:
\begin{align*}
  \exists X_1,\ldots,\exists X_n,\ &\bigl(\forall z_1,\,
  \phi_1(z_1, X_1,\ldots, X_n)\bigr)\\ &\wedge (\exists z_1, \dots,
  \exists z_p,\, \phi_2(z_1 \dots z_p,X_1,\ldots, X_n)\bigr),
\end{align*}
 where
  $\phi_1$ and $\phi_2$ are quantifier-free formulas.
\end{thm}
See \cite[Corollary 4.1]{Thomas} or  \cite[Corollary 4.2]{Thomas2} for a similar result.
{\cR This result is an easy consequence of \cite[Theorem 3.2]{SchBar} (see
  also the corrigendum). We include here a full proof.}
\begin{pf}
	Let $\cal C$ be the set of such formulas.	
	We proceed in three steps:
	\begin{itemize}
        \item Every EMSO-definable set is the projection of a positive combination of
          some $S_{=k}(P)$ and $S_{\geq k}(P)$ (using
          prop. \ref{project} and the preceding corollary)
		\item Every $S_=(P,k)$ (resp. ${S_{\geq}(P,k)}$) is $\cal C$-definable
		\item $\cal C$-definable sets are closed by (finite) union, intersection  and projections.
	\end{itemize}		
        $\cal C$-definable sets are closed by projection using the
        equivalence of prop. \ref{project} in the two directions, the
        note at the end of the proof and some easy formula
        equivalences. The same goes for intersection.
	
	Now we prove that $\cal C$-definable sets are closed by union. The
	difficulty is to ensure that we use only {\cR one universal quantifier}.
	Let $\phi$ and $\phi'$ be two $\cal C$-formulas defining sets $S_1$ and $S_2$.
	We can suppose that $\phi$ and $\phi'$ use the same numbers of second-order
	quantifiers and of first-order existential quantifiers.
	
Then the formula 
{\cR
\begin{align*}
  \exists X, \exists X_1, \dots, \exists X_n,& \forall z_1, \left\{
    \begin{array}{r}
      X(z_1) \iff X(\north(z_1))\\
      X(z_1) \iff X(\east(z_1))\\
      X(z_1) \implies \phi_1(z_1, X_1 \dots X_n)\\
      \neg X(z_1) \implies \phi'_1(z_1, X_1 \dots X_n)\\
    \end{array}
  \right.\\
  &\wedge \exists z_1, \dots, \exists z_p \bigvee\left.
    \begin{array}{r}
      X(z_1) \wedge \phi_2(z_1 \dots z_p, X_1 \dots X_n)\\
      \neg X(z_1) \wedge \phi'_2(z_1 \dots z_p, X_1 \dots X_n)
    \end{array}
  \right.
\end{align*}
}
defines $S_1 \cup S_2$ (the disjunction is obtained through variable
$X$ which is forced to represent either the empty set or the whole
plane $\ZZ^2$).

It is now sufficient to prove that a $S_{=k}(P)$ set (resp. a $S_{\geq k}(P)$
set) is definable by a $\cal C$-formula.
Let $\phi_P(z)$ be the quantifier-free formula such that $\phi_P(z)$ is true if
and only if $P$ appears at position $z$.

Then $S_{=k}(P)$ is definable by
{\cR 
\begin{align*}
  \exists X_1 \dots \exists X_k \exists A_1, \dots,\exists A_k, &\forall x \left\{
    \begin{array}{l}	
      \wedge_i A_i(x) \iff [A_i(\north(x)) \wedge A_i(\east(x))]\\
      \wedge_i X_i(x) \iff \left[A_i(x) \wedge \neg A_i(\south(x)) \wedge \neg A_i(\west(x))\right]\\
      \wedge_{i \not= j} X_i(x) \implies \neg X_j(x)\\
      (\vee_i X_i(x))  \iff \phi_P(x)\\
    \end{array}		  
  \right.\\
  &\wedge \exists z_1, \dots,\exists z_k, X_1(z_1) \wedge\dots \wedge
  X_k(z_k)
\end{align*}
}
{\cR The formula ensures indeed that $A_i$ represents a quarter of the plane,
$X_i$ being a singleton representing the corner of that plane}. If $k = 0$ this becomes $\forall x, \neg \phi_P(x)$.
To obtain a formula for $S_{\geq k}(P)$, change {\cR the last} $\iff$ to a $\implies$ in the formula.
\qed
\end{pf}

% Using some results from \cite{GiamRest}, we will be able in section
% \ref{sec:emso} to obtain a stronger theorem.
%\newpage
\section{Characterization of Subshifts of Finite Type and Sofic Subshifts}
\label{sec:sofic}
\subsection{Subshifts of Finite Type}
We start by a characterization of subshifts of finite type (SFTs, i.e tilings).
The problem with SFTs is that they are closed neither by projection nor by
union\correc{: the 'even shift' is the projection of a SFT but is not itself a SFT (see \cite{lindmarcus}) and if ${{\cal F}_1}=\left\{\tcolor{\cdark}\tcolor{E}\right\}$ and ${{\cal F}_2}=\left\{\tcolor{E}\tcolor{\cdark}\right\}$ then the union $X_{{\cal F}_1}\cup X_{{\cal F}_2}$ is not a SFT.}
As a consequence, the \correc{class of formulas corresponding to SFTs is not
very interesting:}

\begin{thm}
  \label{sft}
	A set of configurations is a SFT if and only if it is defined by a
	formula of the form
    \[\forall z,\, \psi(z)\]
    where $\psi$ is quantifier-free.
\end{thm}
Note that there is only one quantifier in this formula. Formulas with more
than one universal quantifier do not always correspond to SFT: This is due
to SFTs not being closed by union. 
\begin{pf}
 Let $P_1 \dots P_n$ be patterns. To each $P_i$ we associate the
 quantifier-free formula $\phi_{P_i}(z)$ which is true if and only
 if $P_i $ appears at the position $z$.
 Then the subshifts that forbids patterns $P_1 \dots P_n$ is defined by the
 formula:
 \[ \forall z, \neg \phi_{P_1(z)} \wedge \dots \wedge \neg\phi_{P_n(z)} \]
	 
  Conversely, let $\psi$ be a quantifier-free formula.   
  Each term $t_i$ in $\psi$ is of the form $f_i(z)$ where $f_i$ is some combination of
  the functions $\north, \south, \east$ and $\west$, each $f_i$ thus
  representing somehow some vector $z_i$ ($f_i(z) = z+z_i$).
  Let $Z$ be the collection of all vectors $z_i$ that appear in the formula $\psi$.
  Now the fact that $\psi$ is true at the position $z$ only depends on the
  colors of the configurations in points $(z+z_1), \dots, (z+z_n)$, i.e. on the
  \emph{pattern} of domain $Z$ that occurs at position $z$.
  Let $\cal P$ be the set of patterns of domain $Z$ that makes $\psi$ false.  
  Then the set $S$ defined by $\psi$ is the set of configurations where no
  patterns in $\cal P$ occurs, hence a SFT. \qed
\end{pf}	

\subsection{Universal sentences}

Due to the way subshifts are defined, universal quantifiers play an
important role. We now ask the following question: what are the sets
defined by universal formulas?  {\dff First the following lemma shows
  that we can restrict to first-order when considering universal
  formulas.
  \begin{lem}
    Any universal \correc{MSO} formula is equivalent to a first-order universal formula.
  \end{lem}
  \begin{pf}
    A universal formula is equivalent (through permutation of
    universal quantifiers) to a formula of the form $$\forall
    x_1,\ldots,x_p,\forall X_1,\ldots,X_n,\,
    \Phi(X_1,\ldots,X_n,x_1,\ldots,x_p)$$ where $\Phi$ is
    quantifier-free.  Consider the formula 
    $$\psi(X_1,\ldots,X_{n-1},x_1,\ldots, x_p)\equiv\forall X_n,\Phi(X_1,\ldots,X_n,x_1,\ldots,x_p)$$
    Let ${\{t_1,\ldots, t_k\}}$ be the set of terms $t$ such that
    $X_n(t)$ occurs in $\Phi$.  The idea is that the truth value of
    $\Phi(X_1,\ldots,X_n,x_1,\ldots,x_p)$ depends only on the value of
    $X_n$ at positions represented by the $(t_i)$. Depending on
    interpretations of the variables $(x_i)$, interpretations of the
    terms $(t_i)$ may be equal or not.  We say an \correc{assignment}
    ${\rho:\{1,\ldots,k\}\rightarrow\{0,1\}}$ is \emph{sound} if
    ${t_i=t_j\implies \rho(i)=\rho(j)}$. Denote by
    $\phi_\rho(x_1,\ldots,x_p)$ the quantifier-free formula expressing
    this condition:
    $$\phi_\rho(x_1,\ldots,x_p)\equiv\bigwedge_{\{(i,j) : \rho(i)\not=\rho(j)\}}t_j\not=t_j.$$

    Let $\psi_\rho$ denote the formula $\Phi\bigl[X_n(t_i)\leftarrow
    \rho(i)\bigr]$ obtained from $\Phi$ be replacing each occurrence
    of $X_n(t_i)$ by the truth value $\rho(i)$ and this for each
    $i\in\{1,\ldots,k\}$.  For any fixed ${x_1,\ldots,x_p}$, the truth
    value of $\forall X_n\Phi(X_1,\ldots,X_n,x_1,\ldots,x_p)$ is the
    same as the truth value of the \correc{conjunction} of formulas $\psi_\rho$
    for all sound $\rho$.  Hence, we get that
    $\psi(X_1,\ldots,X_{n-1},x_1,\ldots, x_p)$ is equivalent to the
    following quantifier-free formula:
    $$\bigwedge_{\rho:\{1,\ldots,k\}\rightarrow\{0,1\}} \phi_\rho\implies \psi_\rho.$$ 

    We can eliminate this way second order universal quantifiers one
    by one and the lemma follows.\qed
  \end{pf}
For the rest of this section we focus on first-order universal
formulas.} 
The real difficulty is to treat the equality predicate
(=). Without the equality (more precisely if all predicates and
functions are only unary) any first-order universal formula is
equivalent to a conjonction of formulas with only one quantifier and
theorem \ref{sft} applies. The equality predicate intertwines the
variables and makes thing a bit harder to prove.  The reader might for
example try to understand what the following formula exactly means:
\[
	  \forall x,y, \left(\pred{\clime}{x} \wedge \pred{\cblue}{\east(y)}\right)     \implies x = y
\]

To understand it, we will prove an analog of Hanf's
Lemma for universal sentences.

\begin{defn}
	Let $(n,k)$ be integers, and $M,N$ two $Q$-configurations.
	We say that $M \geq_{n,k} N$ if for each $Q$-square pattern $P$ of
	radius less than $n$:
	\begin{itemize}
			  \item If $P$ appears in $M$ exactly $p$ times and $p \leq k$,
				then $P$ appears \correc{at most} $p$ times 		in $N$
		% 	  \item
		% (No condition is required if $P$ appears in $M$ more than $k$ times)
	\end{itemize}			
\end{defn}	
Note that $M$ and $N$ are $(n,k)$ equivalent if and only if $M \geq_{n,k} N$
and $N \geq_{n,k} M$.

\begin{thm}
	For every universal formula $\phi$ there exists $(n,k)$ such that
	if ${M \geq_{n,k} N}$, then $\mathfrak{M} \models\phi \implies
	\mathfrak{N} \models \phi$
\end{thm}
Compare with definition~\ref{hanf:def} and theorem~\ref{hanf:thm}.
Note that Gaifman's Theorem (a more refined version of Hanf's lemma) was
generalized in \cite{GroheWohr} to existential sentences. We may use this
result to obtain ours. %This would however add some unnecessary complications.
\correc{We give below a complete direct proof.}

\begin{pf}
We will translate the usual proof of Hanf's Local Lemma into our special case.
We will try as much as possible to use the same notations as \cite[sec. 2.4]{EF:finmt}.

	We first change the vocabulary and consider that
	$\east, \west, \north, \south$ are binary predicates rather than functions.
    Note that every universal formula will remain a universal formulas,
	albeit with more quantifiers.
	
    Let introduce some notations.
	Let $S(r,a)$ be the set of all points at distance \correc{at most} $r$ of $a$.
	That is $S(r,a) = \{ x: |x-a| \leq r \}$ where $|\cdot|$
	is the Manhattan distance.
	Note that $S(r,a)$ contains $e_r = 2r^2 + 2r+1$ points.	
	Let $S(r,a_1\dots a_p) = \cup_i S(r, a_i)$.
      
   Let $M$ and $N$ be two $Q$-configurations.		 	
    We say that $a_1 \dots a_p \in (\ZZ^2)^p$ and $b_1 \dots b_p \in (\ZZ^2)^p$
	are $k$-isomorphic if
	there exists a bijective map $f$ from $S(3^k, a_1 \dots a_p)$ to 
	${\dff S}(3^k, b_1 \dots b_p)$ that
	preserves the relations, that is
	\begin{itemize}
		\item {\dff $x\ \east\ y \iff f(x)\ \east\ f(y)$}
                \item {\dff $P_c(x)\iff P_c(f(x))$}
		\item $f(a_i) = b_i$.
	  \end{itemize}
	
	It is then clear that if
	$a_1 \dots a_p$ and $b_1 \dots b_p$ are $0$-isomorphic, then we have
	$\mathfrak{M} \models \psi(a_1 \dots a_p) \iff \mathfrak{N} \models
	\psi(b_1 \dots b_p)$ whenever
	$\psi$ is quantifier-free.

	Now take a formula $\phi = \forall x_1 \dots x_n \psi(x_1 \dots x_n)$
	where $\psi$ is quantifier-free.

    Let $M$ and $N$ such that $M \geq_{3^n,n e_{3^n} +1} N$.

	We now prove by induction that
	\begin{center}
	if $a_1 \dots a_p$ and $b_1 \dots b_p$ are $(n-p)$-isomorphic, then for all $b_{p+1}$, there exists $a_{p+1}$ such
	that $a_1 \dots a_{p+1}$ and $b_1 \dots b_{p+1}$ are $(n-p-1)$-isomorphic.
\end{center}
	\begin{itemize}
        \item Case $p = 0$. Let $b_1 \in \ZZ^2$. Consider the pattern
          of radius $3^n$ centered around $b_1$ in $N$.  This pattern
          appears in $N$, hence must appear in $M$ at least one
          time. Take $a_1$ to be the center of this pattern.

\item	Case $p \mapsto p+1$. Let $a_1 \dots a_p$ and $b_1 \dots b_p$ be
  $n-p$ isomorphic. Let $b_{p+1} \in \ZZ^2$.
       \begin{itemize}
		   \item Case 1: $|b_{p+1} - b_i| \leq 2 \times
			 3^{n-p-1}$ for some $b_i$.
			 
			 In this case $S(3^{n-p-1}, b_{p+1}) \subseteq
                         S(3^{n-p},b_i)$.  Hence by taking $a_{p+1} =
                           {\dff f^{-1}(}b_{p+1}{\dff)}$ {\dff where
                           $f$ is the bijective map involved in the
                           ${n-p}$ isomorphism}, it is clear that $a_1
                         \dots a_{p+1}$ and $b_1 \dots b_{p+1}$ are
                         $n-p-1$ isomorphic.
		   \item Case 2: $\forall i, |b_{p+1} - b_i| >  2 \times 3^{n-p-1}$.
		   In this case for every $i$, $S(3^{n-p-1}, b_{p+1}) \cap B(3^{n-p-1}, b_i) =
		   \emptyset$.
			 
			 Consider the pattern $P$ of radius $3^{n-p-1}$ 
			 centered around $b_{p+1}$.
			 
                         This pattern appears {\dff $\alpha$} times
                         inside $S(2\times 3^{n-p-1}, b_1 \dots b_p)$
                         where {\dff $\alpha \leq p e_{2\times
                             3^{n-p-1}}$}.  $P$ appears at least
                         $\alpha+1$ times in $N$ and ${\alpha +1 \leq
                           n e_{3^n} +1}$ hence must appears at least
                         $\alpha+1$ times in $M$. As it appears the
                         same amount of time in $S(2\times 3^{n-p-1},
                         b_1 \dots b_p)$ and $S(2\times 3^{n-p-1},a_1
                         \dots a_p)$ {\dff(by ${n-p}$ isomorphism)},
                         it must appear somewhere else, say centered
                         in $a_{p+1}$. This $a_{p+1}$ is not inside
                         ${\dff S(3^{n-p-1}, a_1 \dots a_p)}$ {\dff
                           because otherwise it would be the center of
                           an occurrence of pattern $P$ inside
                           ${S(2\times 3^{n-p-1},a_1 \dots a_p)}$}. As
                         a consequence, $a_1 \dots a_{p+1}$ and $b_1
                         \dots b_{p+1}$ are $n-p-1$ isomorphic.
	   \end{itemize}

\end{itemize}	
	   
Now suppose that $\mathfrak{M} \models \phi$.
Take $b_1 \dots b_n \in \ZZ^2$.
There exists $a_1 \dots a_n$ such that $a_1 \dots a_n$ and $b_1  \dots b_n$
are $0$-isomorphic. As $\mathfrak{M} \models \phi$ the quantifier-free formula
${\dff\psi}(a_1 \dots a_n)$ is
true in $\mathfrak{M}$. As a consequence ${\dff\psi}(b_1 \dots b_n)$ is true
in  $\mathfrak{N}$. As this is true for all $b_1 \dots b_n$ we obtain
$\mathfrak{N} \models \phi$.
	\qed
\end{pf}	

Given $(P, k)$
we consider the set $S_{\leq k}(P)$ of all configurations such that the pattern
$P$ occurs at most $k$ times ($k$ may be taken equal to $0$)

\begin{cor}
	\label{unisets}
A set is definable by a universal formula if and only if it is a 
positive combination (i.e. unions and intersections) of some $S_{\leq k}(P)$.
\end{cor}	
This corollary should be compared to corollary~\ref{fosets}.
\begin{pf}
Let $\cal C$ be the class of all universal formulas.
It is clear that the set of $\cal C$-defined formulas is closed under
intersection and unions.

Now $S_{\leq k}(P)$ is defined by
\[
	\forall x_1 \dots x_{k+1},
\phi_{P}(x_1) \wedge \dots \wedge \phi_{P}(x_{k+1}) \implies  \bigvee_{i \not=j} x_i = x_j
	\]
For $k = 0$, this becomes $\forall x, \neg\phi_P(x)$.
Hence, every positive combination of some $S_{\leq k}(P)$ is $\cal C$-definable.

Conversely, let $\phi$ be a universal formula and $S$ the set it defines.
Let $(n,k)$ be as in the theorem.

For each configuration $M\in S$ and $P$ a pattern of radius less than \correc{or equal to}
$n$, denote $\phi_M(P) $ the number of times $P$ appears in $M$ with
the convention than ${\phi_M(P) = \infty}$ if $P$ appears more than
$k$ times in $M$.

Consider the set \[S_M = \bigcap_{\stackrel{P|\phi_M(P) \not=\infty,}{{\dff\text{radius}(P)\leq n}}}   S_{\leq
  \phi_M(P)}(P)\]

From the hypothesis on $(n,k)$, we have $S_M \subseteq S$.
It is then easy to see that $S = \cup_M S_M$ where the union is actually
finite (two configurations that are $(n,k)$-equivalent give the same $S_M$).
\qed
\end{pf}
\subsection{Sofic subshifts}

\correc{Recall that sofic subshifts are projections of SFTs.} Using the previous corollary, we are now able to give a characterisation of sofic subshifts:
\begin{thm}
 \label{thm:sofic}
A set $S$ is a sofic subshift if and only if it is definable by a formula of
the form
\[\exists X_1,\exists X_2\ldots, \exists X_n, \forall z_1, \dots, \forall z_p,\,	\psi(X_1,\ldots,X_n,z_1\dots z_p)\]
where $\psi$ is quantifier-free. {\cR Moreover, any such formula is
  equivalent to a formula of the same form but with a single universal
  quantifier ($p=1$)}.
\end{thm}
See \cite{JeTh} for a different proof that eliminates equality predicates one by one.

\begin{pf}
Let $\cal C$ be the clas of all formulas of the form
\[\exists X_1,\ldots, \exists X_n, \forall z \psi(X_1,\ldots,X_n,z)\]
where $\psi$ is quantifier-free. With the help of theorem \ref{sft} and
proposition~\ref{project}, is is quite clear that $\cal C$-defined sets are exactly
sofic subshifts.

Let $\cal D$ be the class of all formulas of the form
\[\exists X_1,\ldots, \exists X_n, \forall z_1 \dots z_p
	\psi(X_1,\ldots,X_n,z_1\dots z_p)\]
where $\psi$ is quantifier-free. 
The previous remark states that sofic subshifts are $\cal D$-defined.

Now we prove that $\cal D$-defined sets are sofic subshifts.
Using (the proof of) proposition~\ref{project}, and the fact that sofic
subshifts are closed under projection, it is sufficient to prove
that universal formulas define sofic subshifts. Using
corollary~\ref{unisets} and the fact that sofic subshifts are closed under
union and projections, it is sufficient to prove that every $S_{\leq k}(P)$ is sofic.
  
Now $S_{\leq k}(P)$ is defined by

\[ \phi:
	\exists S_1 \dots S_k 
\left\{
	  \begin{array}{r}
		  \Psi_i\\
		  \forall x, \vee_i S_i(x) \iff \phi_P(x)
		  \end{array}
	  \right.	  	
\]	

where $\Psi_i$ expresses that $S_i$ has at most one element and is
defined as follows:
\[
                   {\cR \Psi_i\ \overset{def}{=}\ }\exists A, 
                                   \forall x
\left\{
          \begin{array}{r}
                  A(x) \iff A(\north(x)) \wedge A(\east(x))\\
                  S_i(x) \iff A(x) \wedge \neg A(\south(x)) \wedge \neg A(\west(x))     \\
                  \end{array}
\right.                            
                                  \]            
								  
Now with some light rewriting we can transform $\phi$ into a formula of the
class $\cal C$, which proves that $S_{\leq k}(P)$ is $\cal C$-definable,
hence sofic.
\qed

\end{pf}
\section{(E)MSO-definable subshifts}
\label{sec:sep}
\subsection{Separation result}
Theorems~\ref{thm:emso} and \ref{thm:sofic} above suggest that
EMSO-definable subshifts are not necessarily sofic. We will show in
this section that the set of EMSO-definable subshifts is indeed
strictly larger than the set of sofic subshifts.  The proof is based
on the analysis of the computational complexity of forbidden languages
\correc{(the complement of the set of patterns occuring in the considered subshift)}.
%(since we consider only finite patterns, the forbidden language can be
%seen as language of finite words via a standard encoding
%convention)
It is well-known that \correc{any sofic subshift $X$} has a recursively
enumerable forbidden language\correc{: first, with a straightforward backtracking algorithm, we can recursively enumerate all patterns that do not occur in a given SFT $Y$; second, if $X$ is the projection of $Y$, we can recursively enumerate all patterns $P$ such that all patterns $Q$ that projects onto $P$ are forbidden in $Y$}. The following theorem shows that the
forbidden language of an MSO-definable subshift can be arbitrarily
high in the arithmetical hierarchy. 

This is not surprising since
%, on one hand 
arbitrary Turing computation
can be defined via first order formulas (using tilesets)
%, and on the other hand 
and second order quantifiers can be used to simulate quantification of
the arithmetical hierarchy. However, some care must be taken to ensure
that the set of configurations obtained is a subshift.

\begin{thm}
  Let $E$ be an arithmetical set. Then there is an MSO-definable
  subshift with forbidden language $\mathcal F$ such that $E$ reduces
  to $\mathcal F$ (for many-one reduction).
\end{thm}
\begin{pf}[sketch]
  Suppose that the complement of $E$ is defined as the set of integers
  $m$ such that:
  \[\exists x_1,\forall x_2,\ldots, {\cR\exists/}\forall x_n, R(m,x_1,\ldots,x_n)\]
  where $R$ is a recursive relation. We first build a formula $\phi$
  defining the set of configurations representing a successful
  computation of $R$ on some input $m, x_1,\ldots, x_n$.  Consider $3$
  colors $c_l$, $c$ and $c_r$ and additional second order variables
  $X_1,\ldots,X_n$ and $S_1,\ldots, S_n$.  The input
  $(m,x_1,\ldots,x_n)$ to the computation is encoded in unary on an
  horizontal segment using colors $c_l$ and $c_r$ and variables $S_i$
  as separators, precisely: first an occurrence of $c_l$ then $m$
  occurrences of $c$, then an occurrence of $c_r$ and, for each
  successive $1\leq i\leq n$, $x_i$ positions in $X_i$ before a
  position of $S_i$.  Let $\phi_1$ be the FO formula expressing the
  following:
  \begin{enumerate}
  \item there is exactly $1$ occurrence of $c_l$ and the same for
    $c_r$ and all $S_i$ are singletons;
  \item starting from an occurrence $c_l$ and going east until
    reaching $S_n$, the only possible successions of states are those
    forming a valid input as explained above.
  \end{enumerate}
  Now, the computation of $R$ on any input encoded as above can be
  simulated via tiling constraints in the usual way. Consider
  sufficiently many new second order variables $Y_1,\ldots,Y_p$ to
  handle the computation and let $\phi_2$ be the FO formula expressing
  that:
  \begin{enumerate}
  \item a valid computation starts at the north of an occurrence of $c_l$;
  \item there is exactly one occurrence of the halting state
    (represented by some $Y_i$) in the whole configuration.
  \end{enumerate}
  We define $\phi$ by:
  \begin{align*}
    \exists X_1,\forall X_2,\ldots,\exists{\cR/\forall} X_n,\exists
    S_1,\ldots,\exists S_n,\exists Y_1,\ldots,\exists Y_p,
    \phi_1\wedge\phi_2.
  \end{align*}
  Finally let $\psi$ be the following FO formula: ${(\forall z, \neg
    P_{c_l})\vee (\forall z, \neg P_{c_r})}$. Let $X$ be the set
  defined by $\phi\vee\psi$. By construction, a finite
  (unidimensional) pattern of the form ${c_l c^m c_r}$ appears in some
  configuration of $X$ if and only if ${m\not\in E}$. Therefore $E$ is
  many-one reducible to the forbidden language of $X$.

  To conclude the proof it is sufficient to check that $X$ is
  closed. To see this, consider a sequence $(C_n)_n$ of configurations
  of $X$ converging to some configuration $C$. $C$ has at most one
  occurrence of $c_l$ and one occurrence of $c_r$. If one of these two
  states does not occur in $C$ then ${C\in X}$ since $\psi$ is
  verified. If, conversely, both $c_l$ and $c_r$ occur (once each)
  then any pattern containing both occurrences also occurs in some
  configuration $C_n$ verifying $\phi$. But $\phi$ is such that any
  modification outside the segment between $c_l$ and $c_r$ in $C_n$
  does not change the fact that $\phi$ is satisfied provided no new
  $c_l$ and $c_r$ colors are added. Therefore $\phi$ is also satisfied
  by $C$ and $C\in X$.\qed
\end{pf}

The theorem gives the claimed separation result for subshifts of EMSO.

\begin{cor}
  There are EMSO-definable subshifts which are not sofic.
\end{cor}
\begin{pf}
  In the previous theorem, choose $E$, to be the complement of
  the set of integers $m$ for which there is $x$ such that machine $m$
  halts on empty input in less than $x$ steps. $E$ is not recursively
  enumerable and, using the construction of the proof above, it is
  reducible to the forbidden language of an EMSO-definable
  subshift.\qed
\end{pf}

\correc{
\subsection{Subshifts and patterns}
In the previous section we proved that there exists a MSO-definable subshift
for which its forbidden language is not enumerable. This means in
particular that there exists no recursive set $\mathcal F$ of patterns
that defines this subshift, and in particular no \emph{MSO-definable} set of
patterns that defines this subshift. We will show in this section that
this situation does not happen for the classes of subshifts we show
before, that is every subshift of theses classes can be defined
by a set of forbidden patterns of the same (logical) complexity.

For this to work, we now consider a purely relational signature, that
is we consider now $\east, \north, \south, \west$ as binary relations
rather that functions. As we said before, the previous theorems with
the exception of theorem \ref{sft} are still valid in this context.
However with a relational signature, it makes sense to ask whether a given
(finite) pattern $P$ satisfy a formula $\phi$: First-order quantifiers
range over $\mathop{\mathrm{Dom}} P$, the domain of $P$, and second-order monadic quantifiers
over all subsets of $\mathop{\mathrm{Dom}} P$.

\newpage
We now prove

\begin{thm}
	Let $\phi$ be a formula of the form
\[\exists/\forall X_1,\exists/\forall X_2\ldots, \exists/\forall X_n, \forall z_1, \dots, \forall z_p,\,	\psi(X_1,\ldots,X_n,z_1\dots z_p)\]
	
	Then a configuration $M$ satisfies $\phi$ if and only if all
	patterns $P$ of $M$ satisfy $\phi$.
\end{thm}	
\begin{pf}
	The basic idea is to use compactness to bypass the existential
	(second-order) quantifiers.

%We denote by $\mathop{\mathrm{Dom}} P$ the domain of the pattern $P$ and
We denote by $E_{\mathop{\mathrm{Dom}} P}$ the restriction of $E$ to $\mathop{\mathrm{Dom}} P$.
	We prove the following statement by induction: For every subsets $E_1 \dots E_k$ of $\ZZ^d$ and any
	configuration $M$, $(M,E_1, \dots, E_k) \models \phi(X_1 \dots X_k)$ if and only
	if $(P, (E_1)_{\mathop{\mathrm{Dom}} P}, \dots ,(E_k)_{\mathop{\mathrm{Dom}} P}) \models \phi(X_1 \dots X_k)$ for every pattern $P$ of $M$.

    This is clear if $\phi$ has no second-order quantifiers.
	
	Now let $\phi$ be a formula of the previous form.
	Note that it is clear that if $(M,E_1, \dots, E_k) \models
	\phi(X_1 \dots X_k)$ then
	$(P, (E_1)_{\mathop{\mathrm{Dom}} P}, \dots ,(E_k)_{\mathop{\mathrm{Dom}} P}) \models \phi(X_1 \dots X_k)$,
	as the	first order fragment of $\phi$ is universal. We now prove the converse. There are
	two cases:
	\begin{itemize}
		\item First case, $\phi(X_1 \dots X_k) = \forall X \psi(X_1 \dots X_k, X)$.
		  Suppose that $(P,(E_1)_{\mathop{\mathrm{Dom}} P}, \dots, (E_k)_{\mathop{\mathrm{Dom}} P})
		  \models \phi(X_1 \dots X_k)$ for every
		  pattern $P$ of $M$. Let $E$ be a subset of $\ZZ^d$.
		  Now, $(P,(E_1)_{\mathop{\mathrm{Dom}} P}, \dots, (E_k)_{\mathop{\mathrm{Dom}}
			P},E_{\mathop{\mathrm{Dom}} P}) \models \psi(X_1 \dots X_k, X)$ for all
		  patterns $P$ of $M$ by hypothesis, so using the 
		  induction hypothesis, $(M, E_1, \dots , E_k, E) \models \psi(X_1 \dots
		  X_k, X)$ , hence the result
		  $(M, E_1 \dots E_k) \models \forall X \phi(X_1 \dots X_k, X)$.
		\item Second case, $\phi(X_1 \dots X_k) = \exists X \psi(X_1 \dots X_k, X)$.
		  Suppose that $(P,(E_1)_{\mathop{\mathrm{Dom}} P}, \dots, (E_k)_{\mathop{\mathrm{Dom}} P})
		  \models \phi(X_1 \dots X_k)$ for every
		  pattern $P$ of $M$. In particular, for every pattern $P$,
		  there exists a set $E_P$ so that $(P,(E_1)_{\mathop{\mathrm{Dom}} P}, \dots,
		  (E_k)_{\mathop{\mathrm{Dom}} P},E_P)$ satisfies $\psi(X_1, \dots X_k, X)$

		  Let $P_i$ be the pattern of domain $[-i,i]^d$ of $M$, and  $E_{P_i} \subseteq [-i,i]$ the
		  subset given by the previous sentence. We now see $E_{P_i}$
		  as a point in $\{0,1\}^{\ZZ^d}$, and by compactness we know
		  that the set $\{E_{P_i}, i \in \mathbb{N}\}$
		  has an accumulation point $E$. This set $E$ has the
		  following property: for every domain $Z \subseteq \ZZ^d$,
		  there exists $i$ so that $[-i,i]^d$ contains $Z$, and
		  $E_{P_i}$ and $E$ coincide on $Z$.
		  
		  Now we prove that $(M, E_1, \dots E_k, E)$ satisfies $\psi$.
		  Let $P$ be a pattern of $M$. There exists
		  $i$ so that $E_{P_i}$ and $E$ coincide on $\mathop{\mathrm{Dom}} P$.
		  Now by definition of $E_{P_i}$, we have 
		  $(P_i,(E_1)_{\mathop{\mathrm{Dom}} P_i}, \dots, (E_k)_{\mathop{\mathrm{Dom}} P_i},E_{P_i}) \models
		  \psi(X_1\dots X_k, X)$.
		  However, as $P$ is a subpattern of $P_i$, and the fact that
		  the first order fragment of $\psi$ is universal, we have that
		  $(P,(E_1)_{\mathop{\mathrm{Dom}} P}, \dots, (E_k)_{\mathop{\mathrm{Dom}} P},(E_{P_i})_{\mathop{\mathrm{Dom}} P}) \models
		  \psi(X_1\dots X_k, X)$. Now $E$ coincide with $E_{P_i}$ on
		  $\mathop{\mathrm{Dom}} P$, so that we have $(P,(E_1)_{\mathop{\mathrm{Dom}} P}, \dots, (E_k)_{\mathop{\mathrm{Dom}} P},E_{\mathop{\mathrm{Dom}}
			P}\models \psi(X_1 \dots X_k, X)$). Using the induction hypothesis, 
		  we have proven that
		  $(M, E_1, \dots E_k, E) \models \psi(X_1 \dots X_k, X)$, hence 
		  $(M,E_1 \dots E_k \models \exists X \psi(X_1 \dots X_k, X)$.
	\end{itemize}
	\qed
\end{pf}	
\begin{cor}
	If $S$ is a subshift defined by a formula $\phi$ of the form of
	the preceding theorem, then $S = X_{\cal F}$ where $\cal F$ is the
	set of words that do not satisfy $\phi$.		
\end{cor}	
In particular, in dimension $1$, if a subshift is defined by a EMSO
formula (is sofic), then it is defined by a EMSO-definable set of
forbidden words, ie a regular set. Similarly, if a subshift is
defined by a (universal) FO formula, it is defined by a (universal)
FO-definable set of forbidden words, hence in particular by a strongly
threshold locally testable language \cite{BeauquierPin} (compare with
corollary \ref{unisets}).

The previous section shows that the corollary does not work for
arbitrary formula $\phi$. Indeed, for any MSO-formula $\phi$, the set
of words that do not satisfy $\phi$ is recursive, but there exists
MSO-definable subshifts that cannot be given by a recursive set of
forbidden words.
}

\subsection{Definability of MSO-subshifts}
As we saw before, sets defined by MSO-formulas are not always subshifts.
We will try in this section to find a fragment of MSO that contains only
subshifts and contain all of them. This fragment is somewhat ad hoc. Finding
a more reasonable fragment is an interesting open question.

We first begin by a definition
\begin{defn}
\[
	fin(S):  \exists A, \exists B
\left\{
	  \begin{array}{r}
                  \forall x, A(x) \iff A(\north(x)) \wedge A(\east(x))\\
                  \forall x, B(x) \iff A(\south(x)) \wedge A(\west(x))\\
				  \exists x, A(x) \wedge \neg A(\south(x)) \wedge \neg A(\west(x))\\
				  \exists x, B(x) \wedge \neg B(\north(x)) \wedge \neg B(\east(x))\\
				  \forall x, S(x) \implies A(x) \wedge B(x)
				  \end{array}
\right.
\]
\end{defn}	
It is easy to prove that $fin(S)$ is true if and only if $S$ is finite
(there are finitely many $x$ such that $S(x))$. Indeed $A$ and $B$
represent quarter of planes, and $S$ must be contained in the square
delimited by the two quarter of planes.
Any other formula true only if $S$ is finite would work in the following

\begin{thm}
Let $\correc{X}$ be a MSO-definable set.
Then $\correc{X}$ is a subshift if and only if it is definable by a formula of the form

\[
	\forall S, fin(S)  \implies  \exists B_1 \dots B_k, \psi(S, B_1 \dots B_k) \wedge \forall x_1 \dots x_n S(x_1) \wedge
	\dots S(x_p) \implies \theta(S,B_1 \dots B_k,x_1 \dots x_p)
	\]
	where
\begin{itemize}
	\item $\psi$  is any MSO-formula not containing the predicates $P_c$.
	\item $\theta$ is quantifier-free.
\end{itemize}	
\end{thm}
Note that this formula can be written more concisely as
\[
	\forall^{fin} S, \exists \overline{B} \psi(S, \overline{B}) \wedge 
	\forall \overline{x} \in S^p,   \theta(S,\overline{B}, \overline{x})
	\]
\begin{pf}
	
	First we prove that such a formula $\phi$ defines a subshift $X$ .
	For this, we prove that the set $X$ is closed.
Consider a sequence $M_1 \dots M_n \dots$ of configurations of $X$ converging to
some configuration $M$. We must prove that $M \in X$.

Let $S$ be a finite set. 
Now consider the formula $\theta$. As it is quantifier-free, it is local:
the value of $\theta(S,B_1 \dots B_k,x_1 \dots x_n)$ depends only of what
happens around $x_1 \dots x_n$. As each $x_1 \dots x_n$ must be in $S$,
there exists a finite $S' \supset S$ such that the value of
$\forall x_1 \in S \dots x_n\in S, \theta(S,B_1 \dots B_k,x_1 \dots x_n)$ depends only
of the value of the predicates $S, P_c$ and $B_i$ on $S'$.

Now $M_i$ converges to $M$. This means that there exists $p$ such that $M_p$
and $M$ coincides on $S'$.
For this $M_p$, there exists some $B_1 \dots B_k$ such that
we have 
${\mathfrak{M}_p \models \psi(S, B_1 \dots B_k) \wedge \forall x_1 \in S \dots \forall
	x_p\in S, \theta(S,B_1 \dots B_k,x_1 \dots x_n)}$.
	Then this formula is also true on $\mathfrak{M}$ (Note indeed that
	$\psi(S, B_1 \dots B_k)$ does not depend on the configuration).
	
Hence we have found for every $S$ some $B_i$ that makes the formula true,
that is we have proven $\mathfrak{M} \models \phi$. Therefore $X$ is closed, hence a subshift.

Now let $X$ be a MSO-definable subshift. $X$ is defined by a formula $\phi$.
Change each $P_c$ in $\phi$ by a predicate $B_c$ to obtain $\psi_1$.
Define \[
\psi(\overline{B}) = \forall x \left(\bigvee_c B_c(x)\right) \wedge
\left(
\bigwedge_{c\not=c'} \neg (B_c(x) \wedge B_{c'}(x))\right) \wedge \psi_1(\overline{B})\]

Then $X$ is defined by 
\[
\phi:	\forall^{fin} S, \exists \overline{B} \psi(\overline{B}) \wedge	
	\forall x \in S,	
	\bigwedge_c \left(B_c(x) \iff P_c(x)\right)	
	\]

Indeed $M$ satisfies $\phi$ and only if every pattern of ${\dff M}$ is a pattern
in some configuration of $X$. \qed

\end{pf}
\section{A Characterization of EMSO}
\label{sec:emso}

EMSO-definable sets are projections of FO-definable sets
(proposition~\ref{project}). Besides, sofic subshifts are projections
of subshifts of finite type (or tilings). Previous results show that
the correspondence sofic$\leftrightarrow$EMSO fails. However, we will
show in this section how EMSO can be characterized through projections
of ``locally checkable'' configurations.

Corollary~\ref{fosets} expresses that FO-definable sets are
essentially captured by counting occurrences of patterns up to some
value. The key idea in the following is that this counting can be
achieved by local checkings (equivalently, by tiling constraints),
provided it is limited to a finite and explicitly delimited
region. This idea was successfully used in~\cite{GiamRest} in the
context of picture languages: pictures are rectangular finite patterns
with a border made explicit using a special state (which occurs all
along the border and nowhere else).  
%In the following lemma, we
%reformulate the result obtained in~\cite{GiamRest} in a more general
%way adapted to our setting. 
We will proceed here quite differently. Instead of putting special states on
borders of some rectangular zone, we will simply require that two
special subsets of states $Q_0$ and $Q_1$ are present in the
configuration: we call a \emph{$(Q_0,Q_1)$-marked configuration} any
configuration that contains both a color $q\in Q_0$ and some
color $q'\in Q_1$ somewhere. By extension, given a subshift $\Sigma$ over
$Q$ and two subsets ${Q_0\subseteq Q}$ and ${Q_1\subseteq Q}$, the
\emph{doubly-marked set} $\Sigma_{Q_0,Q_1}$ is the set of
$(Q_0,Q_1)$-marked configurations of $\Sigma$. Finally, a
\emph{doubly-marked set of finite type} is a set $\Sigma_{Q_0,Q_1}$
for some SFT $\Sigma$ and some $Q_0,Q_1$.

\begin{lem}
  \label{lem:count}
  Consider any finite pattern $P$ and any $k\geq 0$. Then $S_{=k}(P)$ is the
  projection of some doubly-marked set of finite type. The same result
  holds for $S_{\geq k}(P)$. 

  Moreover, any positive combination (union and intersection) of
  projections of doubly-marked sets of finite type is also the
  projection of some doubly-marked sets of finite type.
\end{lem}

\begin{pf}[sketch]
  \correc{For the first part of the theorem statement,} we consider some base alphabet $Q$, some pattern $P$ and some $k\geq
  0$. We will build a doubly-marked set of finite type over alphabet
  ${Q'=Q\times Q_+}$ and then project back onto $Q$.  \correc{The set} $Q_+$ is itself a
  product of different layers. The first layer can take values
  $\{0,1,2\}$ and is devoted to the definition of the marker subsets
  $Q_0$ and $Q_1$: a state is in $Q_i$ for ${i\in\{0,1\}}$ if and only
  if its value on the first layer is $i$.

  We first show how to convert the appearance in a configuration of
  two marked positions, by $Q_0$ and $Q_1$, into a locally
  identifiable rectangular zone. The zone is defined by two opposite
  corners corresponding to an occurrence of some state of $Q_0$ and
  $Q_1$ respectively. This can be done using only finite type
  constraints as follows.  By adding a new layer of states, one can
  ensure that there is a unique occurrence of a state of $Q_0$ and
  maintain everywhere the following information:
  \begin{enumerate}
  \item $N_{Q_0}(z)\equiv$ the position $z$ is at the north of the (unique)	occurrence
    of a state from $Q_0$,
  \item $E_{Q_0}(z)\equiv$ the position $z$ is at the east of the occurrence
    of a state from $Q_0$.
  \end{enumerate}
  The same can be done for $Q_1$. From that, the membership to the
  rectangular zone is defined at any position $z$ by the following
  predicate (see figure~\ref{fig:lemcount}):
  \[Z(z)\equiv N_{Q_0}(z)\not=N_{Q_1}(z)\wedge
  E_{Q_0}(z)\not=E_{Q_1}(z).\] 

  \begin{figure}
    \label{fig:lemcount}
    \centering
    \begin{tikzpicture}
      \fill[fill=gray!10] (-.5,-2)--(-.5,2)--(.5,2)--(.5,-2);
      \fill[fill=gray!50] (-3,.5)--(3,.5)--(3,-.5)--(-3,-.5);
      \fill[fill=gray] (-.5,-.5)--(-.5,.5)--(.5,.5)--(.5,-.5);      
      \draw (-3,.5)--(3,.5);
      \draw (-3,-.5)--(3,-.5);
      \draw (-.5,2)--(-.5,-2);
      \draw (.5,2)--(.5,-2);
      \draw[->,>=triangle 45] (1.5,2) node[right] {$E_{Q_0}(z)\not=E_{Q_1}(z)$} -- (0,1);
      \draw[->,>=triangle 45] (-2.5,1.5) node[above] {$N_{Q_0}(z)\not=N_{Q_1}(z)$} -- (-1.5,0);
      \draw (.5,.5) node[above right] {$Q_0$};
      \draw (-.5,-.5) node[below left] {$Q_1$};
      \fill (.5,.5) circle (2pt);
      \fill (-.5,-.5) circle (2pt);
    \end{tikzpicture}
    \caption{The rectangular zone in dark gray defined by predicate
      $Z(z)$.}
  \end{figure}
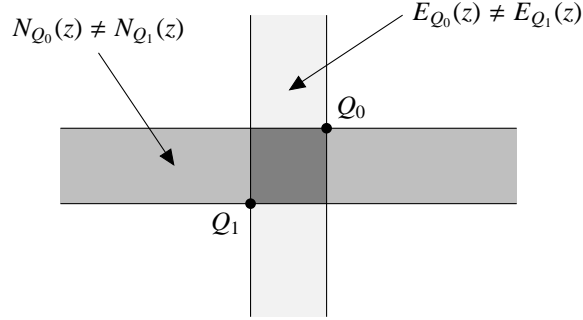
  We can also define locally the border of the zone: precisely, cells
  not in the zone but adjacent to it.  Now define $P(z)$ to be true if
  and only if $z$ is the lower-left position in an occurrence of the
  pattern $P$. We add $k$ new layers, each one storing (among other
  things) a predicate $C_i(z)$ verifying
  {\cR \[C_i(z) \Rightarrow Z(z)\wedge P(z) \wedge \bigwedge_{j\not=i}
  \neg C_j(z).\]} Moreover, on each layer $i$, we enforce that exactly $1$
  position $z$ verifies $C_i(z)$: this can be done by maintaining
  north/south and east/west tags (as for $Q_0$ above) and requiring
  that the north (resp. south) border of the rectangular zone sees
  only the north (resp. south) tag and the same for
  east/west. Finally, we add the constraint:
  \[P(z)\wedge Z(z)\Rightarrow \bigvee_{i} C_i\] expressing that each
  occurrence of $P$ in the zone mut be ``marked'' by some $C_i$.
  Hence, the only admissible $(Q_0,Q_1)$-marked configurations are
  those whose rectangular zone contains exactly $k$ occurrences of
  pattern $P$. We thus obtain exactly $S_{\geq k}(P)$ after
  projection \correc{onto $Q$}. To obtain $S_{= k}(P)$, it suffices to add the
  constraint:
  \[P(z)\Rightarrow Z(z)\] in order to forbid occurrences of $P$
  outside the rectangular zone. 

  To conclude the proof we show that finite unions or intersections of
  projections of doubly-marked sets of finite type are also
  projections of doubly-marked sets of finite type.  Consider two SFT
  $X$ over $Q$ and $Y$ over $Q'$ and two pairs of marker subsets
  $Q_0,Q_1\subseteq Q$ and $Q_0',Q_1'\subseteq Q'$. Let
  ${\pi_X:Q\rightarrow A}$ and ${\pi_Y:Q'\rightarrow A}$ be two
  projections. \correc{Denote by $\Sigma_X$ and $\Sigma_Y$ (resp.) the subsets of $A^{\ZZ^2}$ defined by ${\pi_X\bigl(X_{Q_0,Q_1}\bigr)}$ and ${\pi_Y\bigl(Y_{Q'_0,Q'_1}\bigr)}$. We want to show that both the union $\Sigma_x\cup\Sigma_Y$ and the intersection $\Sigma_X\cap\Sigma_Y$ are projections of some doubly marked sets of finite type.}

  First, for the case of union, we can suppose (up to renaming of
  states) that $Q$ and $Q'$ are disjoint and define the SFT $\Sigma$
  over alphabet $Q\cup Q'$ as follows:
  \begin{itemize}
  \item 2 adjacent positions must be both in $Q$ or both in $Q'$;
  \item any pattern forbidden in $X$ or $Y$ is forbidden in $\Sigma$.
  \end{itemize}
  Clearly, ${\pi(\Sigma_{Q_0\cup Q_0',Q_1\cup Q_1'}) =
    \pi_X(X_{Q_0,Q_1})\cup\pi_Y(Y_{Q_0',Q_1'})}$ where $\pi(q)$ is
  $\pi_X(q)$ when $q\in Q$ and $\pi_Y(q)$ else.
  
  Now, for intersections, consider the SFT $\Sigma$ over
  the fiber product
  
  \[{Q_\times=\{ (q,q') \in Q\times Q' | \pi_X(q) =
	\pi_Y(q')\}}\] and defined as follows:
 a pattern is forbidden if its projection on the component
    $Q$ (resp. $Q'$) is forbidden in $X$ (resp. $Y$);

  If we define $\pi$ as $\pi_X$ applied to the $Q$-component of
  states, and if $E$ is the set of configuration of $\Sigma$ such that
  states from $Q_0$ and $Q_1$ appear on the first component and states
  from $Q_0'$ and $Q_1'$ appear on the second one, then we have:
  \[\pi(E) = \pi_X(X_{Q_0,Q_1})\cup\pi_Y(Y_{Q_0',Q_1'}).\]
  To conclude the proof, it is sufficient to obtain $E$ as the
  projection of some doubly-marked set of finite type. This can be
  done starting from $\Sigma$ and adding a new component of states
  whose behaviour is to define a zone from two markers (as in the
  first part of this proof) and check that the zone contains
  occurrences of $Q_0$, $Q_1$, $Q_0'$ and $Q_1'$ in the appropriate
  components.\qed
\end{pf}

\begin{thm}
  A set is EMSO-definable if and only if it is the projection of a
  doubly-marked set of finite type.
\end{thm}
\begin{pf}
  First, a doubly-marked set of finite type is an FO-definable set
  because SFT are FO-definable (theorem~\ref{sft}) and the restriction
  to doubly-marked configurations can be expressed through a simple
  existential FO
  formula. Thus the projection of a doubly-marked set of finite type
  is EMSO-definable.

  The opposite direction follows immediately from
  proposition~\ref{project} and corollary~\ref{fosets} and the lemma
  above.\qed
\end{pf}

At this point, one could wonder whether considering simply-marked set
of finite type is sufficient to capture EMSO via projections. In fact
the presence of $2$ markers is necessary in the above theorem:
considering the set $\Sigma_{Q_0,Q_1}$ where $\Sigma$ is the full
shift $Q^{\ZZ^2}$ and $Q_0$ and $Q_1$ are distinct singleton subsets
of $Q$, a simple compactness argument allows to show that it is not
the projection of any simply-marked set of finite type.

\section{Open Problems}
\begin{itemize}
\item Is the second order alternation hierarchy strict for MSO
  (considering our model-theoretic equivalence)?
\item One can prove that theorem~\ref{sft} also holds for formulas
  of the form:
    \[\forall X_1 \dots \forall X_n, \forall z,\, \psi(z, X_1\dots X_n)\]
    where $\psi$ is quantifier-free.  Hence, adding universal
    second-order quantifiers does not increase the expression power of
    formulas of theorem~\ref{sft}. More generally, let $\mathcal C$ be the class of formulas
    of the form
  \[\forall X_1,\exists X_2,\ldots, {\cR\forall/}\exists X_n, \forall z_1,\ldots,\forall z_p,\phi(X_1,\ldots,X_n,z_1,\ldots,z_p).\]
  One can check that any formula in $\mathcal C$ defines a
  subshift. Is the second-order quantifiers alternation hierarchy
  strict in $\mathcal C$? On the contrary, do all formulas in $\mathcal C$
  represent sofic subshifts ?
\end{itemize}

\bibliographystyle{elsarticle-num}
\bibliography{article-journal}

\begin{thebibliography}{10}
\expandafter\ifx\csname url\endcsname\relax
  \def\url#1{\texttt{#1}}\fi
\expandafter\ifx\csname urlprefix\endcsname\relax\def\urlprefix{URL }\fi
\expandafter\ifx\csname href\endcsname\relax
  \def\href#1#2{#2} \def\path#1{#1}\fi

\bibitem{Thomas}
W.~Thomas, Handbook of Formal Languages, Vol. 3. Beyond Words, Springer, 1997,
  Ch. {Languages, Automata, and Logic}.

\bibitem{Matz}
O.~Matz, N.~Schweikardt, Expressive power of monadic logics on words, trees,
  pictures, and graphs, in: E.~G. J.~Flum, T.~Wilke (Eds.), Logic and Automata:
  History and Perspectives, Texts in Logic and Games, Amsterdam University
  Press, 2007, pp. 531--552.

\bibitem{D11}
M.~Delacourt, Rice's theorem for mu-limit sets of cellular automata, in: ICALP,
  2011, to appear.

\bibitem{BoyerT10}
L.~Boyer, G.~Theyssier, On factor universality in symbolic spaces, in: MFCS,
  2010, pp. 209--220.

\bibitem{BoyerT09}
L.~Boyer, G.~Theyssier, On local symmetries and universality in cellular
  automata, in: STACS, 2009, pp. 195--206.

\bibitem{LafitteW07}
G.~Lafitte, M.~Weiss, Universal tilings, in: STACS, 2007, pp. 367--380.

\bibitem{LafitteW08}
G.~Lafitte, M.~Weiss, Computability of tilings, in: IFIP TCS, 2008, pp.
  187--201.

\bibitem{DotyPRSS10}
D.~Doty, M.~J. Patitz, D.~Reishus, R.~T. Schweller, S.~M. Summers, Strong
  fault-tolerance for self-assembly with fuzzy temperature, in: FOCS, 2010, pp.
  417--426.

\bibitem{DotyLPSW10}
D.~Doty, J.~H. Lutz, M.~J. Patitz, S.~M. Summers, D.~Woods, Intrinsic
  universality in self-assembly, in: STACS, 2010, pp. 275--286.

\bibitem{wangpatternrecoII}
H.~Wang, Proving theorems by pattern recognition ii, Bell system technical
  journal 40 (1961) 1--41.

\bibitem{Ber-undecidability-dp}
R.~Berger, The undecidability of the domino problem, Memoirs American
  Mathematical Society 66 (1966) 1966.

\bibitem{classicaldecisionproblem}
E.~B\"orger, E.~Gr\"adel, Y.~Gurevich, The classical decision problem,
  Springer-Verlag Telos, 1996.

\bibitem{Kuske}
D.~Kuske, M.~Lohrey, {Logical aspects of Cayley-graphs: the group case}, Annals
  of Pure and Applied Logic 131~(1-3) (2005) 263--286.

\bibitem{Seese}
D.~Seese, {The structure of the models of decidable monadic theories of
  graphs}, Annals of pure and applied logic 53~(2) (1991) 169--195.

\bibitem{makowsky74}
J.~Makowsky, On some conjectures connected with complete sentences, Fund. Math.
  81 (1974) 193--202.

\bibitem{poizat80}
B.~Poizat, Une th\'eorie finiement axiomatisable et superstable, Groupe
  d'\'etude des th\'eories stables 3~(1) (1980-1982) 1--9.

\bibitem{Robinson}
R.~M. Robinson, {Undecidability and Nonperiodicity for Tilings of the Plane},
  Inventiones Math. 12.

\bibitem{oger04}
F.~Oger, Algebraic and model-theoretic properties of tilings, Theoretical
  computer science 319 (2004) 103--126.

\bibitem{jac:modth}
A.~Ballier, E.~Jeandel, Tilings and model theory, First Symposium on Cellular
  Automata Journ\'ees Automates Cellulaires.

\bibitem{GiamRest2}
D.~Giammarresi, A.~Restivo, Handbook of Formal Languages, Vol. 3. Beyond Words,
  Springer, 1997, Ch. Two-Dimensional Languages.

\bibitem{GiamRest}
D.~Giammarresi, A.~Restivo, S.~Seibert, W.~Thomas, {Monadic Second-Order Logic
  over Rectangular Pictures and Recognizability by Tiling Systems}, Information
  and Computation 125 (1996) 32--45.

\bibitem{Weiss}
B.~Weiss, Subshifts of finite type and sofic systems, Monatshefte f{\"u}r
  Mathematik 77 (1973) 462--474.

\bibitem{AnselmoJM09}
M.~Anselmo, N.~Jonoska, M.~Madonia, {Framed Versus Unframed Two-Dimensional
  Languages}, in: 35th Conference on Current Trends in Theory and Practice of
  Computer Science, SOFSEM, 2009, pp. 79--92.

\bibitem{Alten}
J.-H. Altenbernd, W.~Thomas, S.~Wohrle, Tiling systems over infinite pictures
  and their acceptance conditions, in: Developments in Language Theory, 2003.

\bibitem{Harel}
D.~Harel, {Recurring Dominoes: Making the Highly Undecidable Highly
  Understandable}, Annals of Discrete Mathematics 24 (1985) 51--72.

\bibitem{JeTh}
E.~Jeandel, G.~Theyssier, {Subshifts, Languages and Logic}, in: Developments in
  Language Theory (DLT), no. 5583 in LNCS, Springer, 2009, pp. 288--299.

\bibitem{hanf65}
W.~Hanf, {Model-theoretic methods in the study of elementary logic}, in:
  Symposium in the Theory of Models, 1965, pp. 132--145.

\bibitem{EF:finmt}
H.-D. Ebbinghaus, J.~Flum, Finite Model Theory, Springer Monographs in
  Mathematics, Springer, Berlin, 1995.

\bibitem{Libkin}
L.~Libkin, {Elements of Finite Model Theory}, Texts in Theoretical Computer
  Science, an EATCS Series, Springer Verlag, 2004.

\bibitem{Thomas2}
W.~Thomas, {On logics, tilings, and automata}, in: S.~Berlin (Ed.), Proceedings
  of the 18th ICALP, Vol. 510 of LNCS, 1991, pp. 441--454.

\bibitem{SchBar}
T.~Schwentick, K.~Barthelmann, {Local Normal Forms for First-Order Logic with
  Applications to Games and Automata}, Discrete Mathematics and Theoretical
  Computer Science 3 (1999) 109--124.

\bibitem{lindmarcus}
D.~Lind, B.~Marcus, An introduction to symbolic dynamics and coding, Cambridge
  University Press, 1995.

\bibitem{GroheWohr}
M.~Grohe, S.~W\"ohrle, {An existential locality theorem}, Annals of Pure and
  Applied Logic 129~(1-3) (2004) 131--148.

\bibitem{BeauquierPin}
D.~Beauquier, J.-E. Pin, {Languages and scanners}, Theoretical Computer Science
  84 (1991) 3--21.

\end{thebibliography}

\end{document}